# The Howard-Harvard effect: Institutional reproduction of intersectional inequalities

**Authors:** Diego Kozlowski[1,2,3], Thema Monroe-White[4], Vincent Larivière[1,2,5,6], Cassidy R. Sugimoto[2,5]*

[1] École de bibliothéconomie et des sciences de l'information, Université de Montréal, Montréal, QC, Canada.

[2] School of Public Policy, Georgia Institute of Technology

[3] Faculty of Science, Technology and Medicine, University of Luxembourg, Esch-Sur-Alzette 4364, Luxembourg

[4] Department of Technology, Entrepreneurship and Data Analytics, Berry College, Mount Berry, GA, US

[5] Department of Science and Innovation-National Research Foundation Centre of Excellence in Scientometrics and Science, Technology and Innovation Policy, Stellenbosch University, Stellenbosch 7602, South Africa.

[6] Observatoire des sciences et des technologies, Centre interuniversitaire de recherche sur la science et la technologie, Université du Québec à Montréal, Montréal, QC, Canada.

*Corresponding author.

**Email:** sugimoto@gatech.edu

Conceptualization: DK, VL, CRS, TMW
Data curation: DK, VL
Formal Analysis: DK
Funding acquisition: CRS, TMW, DK, VL
Investigation: DK, VL
Methodology: DK, VL, CRS, TMW
Project administration: VL, CRS, TMW
Resources: VL, DK
Software: DK
Supervision: VL, CRS, TMW
Validation: DK, VL
Visualization: DK
Writing – original draft: DK, VL, CRS, TMW
Writing – review & editing: DK, VL, CRS, TMW

**Competing Interest Statement:** Authors declare that they have no competing interests.

**Classification:** Social, Social Sciences.

**Abstract**

The US higher education system concentrates the production of science and scientists within a few institutions. This has implications for minoritized scholars and the topics with which they are disproportionately associated. This paper examines topical alignment between institutions and authors of varying intersectional identities, and the relationship with prestige and scientific impact. We observe a Howard-Harvard effect, in which the topical profile of minoritized scholars are amplified in mission-driven institutions and decreased in prestigious institutions. Results demonstrate a consistent pattern of inequality in topics and research impact. Specifically, we observe statistically significant differences between minoritized scholars and White men in citations and journal impact. The aggregate research profile of prestigious US universities is highly correlated with the research profile of White men, and highly negatively correlated with the research profile of minoritized women. Furthermore, authors affiliated with more prestigious institutions are associated with increasing inequalities in both citations and journal impact. Academic institutions and funders are called to create policies to mitigate the systemic barriers that prevent the United States from achieving a fully robust scientific ecosystem.

# Introduction

Race and gender disparities in the research workforce affect what type of research is produced and its relevance for society (Kozlowski, Larivière, et al., 2022a). These intersectional disparities are persistent and pervasive: Black or African American, Hispanic or Latino, and American Indian or Alaska Native students and women generally, account for fewer earned doctorates (NSF, 2021b) and produce fewer scientific articles (Kozlowski, Larivière, et al., 2022a; Larivière et al., 2013) than would be expected given their representation in the population. Disparities are amplified across the research pipeline: the share of academic positions held by minoritized scholars is less than 9%— a percentage that is considerably less than their share of doctoral graduates (NSF, 2021b). Barriers to entry and participation in science can be seen as consequences of inequalities in peer review in journals (Erosheva et al., 2020; E. Ross, 2017) and funding applications (Chen et al., 2022; Ginther et al., 2011). Once published, the work of minoritized scholars receives less visibility in the media (Peng et al., 2022) and fewer citations . These disparities are compounded at the intersection of race and gender identities of authors (Crenshaw, 1991), and mediated by research topics (Bertolero et al., 2020; Kozlowski, Larivière, et al., 2022a).

Universities play a key role in creating policies and practices that shape the social structure in which research is conducted. In the United States (US), there is considerable heterogeneity across institutions in terms of history, mission, and resources, with implications for the composition of the faculty, staff, and students. This is particularly the case in mission-driven institutions, such as Historically Black Colleges and Universities (HBCUs) and Women's College (WCs) (McGee et al., 2021; Sax et al., 2014), which focus recruitment on specific populations. These specific orientations have significant implications for the diversity of the scientific workforce. For example, 23% of Black and African American students who earned a doctorate degree in science and engineering between 2015 and 2019 received a bachelor's degree from an HBCU (NSF, 2021b; Owens et al., 2012).

The US higher education system is an extremely stratified environment, with sharp inequalities in access. For example, graduates from the most "prestigious" 20% of US universities occupy 80% of all faculty positions in the country (Wapman et al., 2022), with universities rarely hiring graduates from lower-ranked institutions (Clauset et al., 2015). Faculty at prestigious institutions tend to accumulate other benefits, such as increased funding and access to larger doctoral student labor markets (Zhang et al., 2022). These benefits lead to higher productivity and recognition (Way et al., 2019), which reinforces hiring inequalities, particularly for women (Clauset et al., 2015; LaBerge et al., 2022). The scientific consequences are important: scientific ideas spread more quickly and with greater impact when they come from prestigious institutions (Morgan et al., 2018). Knowledge generation, dissemination, and human capacity development are strongly concentrated among a few institutions, with implications for the research portfolio of the nation, and the degree to which science is serving all of society.

In this paper, we analyze how institutional prestige relates to the race and gender identity of authors, research topics, and scientific impact. We examine three different levels of institutional prestige: *perceived* prestige, drawn from the US News & World Report institutions rankings; *research* prestige, measured as the institution's average of field-normalized citations; and *selectivity* prestige, using Carnegie's Selectivity Index which measures acceptance rates of undergraduate students. We also analyze Historically Black Colleges and Universities (HBCU), Women's Colleges (WC), and Hispanic Serving Institutions (HSI), as exemplars of mission-driven and community-serving institutions. Two main questions are addressed: (1) how do institutions of varying prestige differ in topical orientation and how does this relate to intersectional identities of authors? and, (2) which is the relation between impact and institutional prestige?

## Methods

To answer these questions, we leveraged a dataset of more than 4.5 million articles published between 2008 and 2020, indexed in the Web of Science (WOS), and affiliated with 685 US universities. Following the method developed by Kozlowski, Murray et al. (2022), first authors of the selected papers were assigned a probability over each racial group based on the association between their family names and racial categories found in the 2010 US Census (USBC, 2016) (See more detail on SI Data & Methods). Gender was inferred in a binary fashion using authors' given names, based on Larivière et al. (2013). Subsequently, we consider an author's identity as the combination of four racial categories—Black, Latinx[1], Asian, and White—and a binary gender indicator. Given the limitations of the data and inference algorithms, we were unable to robustly assign distributional properties for Native American and "Two or more races", nor were we able to code beyond a binary operationalization of gender. We acknowledge the complex history of the U.S. Census classifications of race (Zuberi, 2000) and the assumption of within group homogeneity that is implied (e.g., Black and African American). Given the assumption of equivalence between the U.S. Census population and the Web of Science population, our method potentially overestimates the proportion of Black and Latinx authors (LaBerge et al., 2022). Furthermore, we acknowledge that there are several additional and important dimensions of intersectionality, such as class, sexual orientation, disability, immigration/citizenship status, and language. In addition to other minoritized and marginalized identities that we do not analyze in this study. These limitations highlight the importance of triangulation and comparison with studies based on surveys and author self-identification (Langin, 2020). While self-identification of identity variables is always preferred, large-scale bibliometric databases do not provide such information. Therefore, we used the best possible approximation for large-scale analysis that complement—and are complemented by–in-depth case survey studies of a specific subgroup of the population. We strongly encourage the use of self-identified and disaggregated data when available.

---

[1] Black and African American are considered as a single category and termed "Black" in this paper; Latinos are referred to as Latinx. We acknowledge and consider the complexities of this aggregation in the Discussion.

Likewise, given that our unit of analysis is articles, and we assign the first author's inferred race and gender as the identity associated with the entire article, we overrepresent dominant junior authorships (Larivière et al., 2016).. Thus, the combination of the potential overestimation of Black and Latinx authors and sole consideration of first authors suggests an optimistic view, as disparities are more extreme for last (i.e., more senior) authors (see Figure S1). However, although White men are more overrepresented as last authors than first authors, in our previous work we found strong homophily patterns on co-authorship for this same dataset (Kozlowski, Larivière, et al., 2022b) (see Fig. S2), which implies that similar results are expected to be found when considering all authors.

We use historical WoS data to compute the average of field-normalized citations of US universities between 1980 and 2019. We consider three groups of prestige for the US News & World Report: Top 10, Top 100 (excluding the Top 10), and institutions ranked below 100. Institutional prestige is a multidimensional concept, and therefore no single metric can fully capture it. We complement the *perceived* prestige from US News & World Report with two alternative measures. The average citation rank divides institutions into three equally sized groups as a function of the mean impact of their research articles (we also consider a deciles version and a continuous version for the linear regression models in the SI), while the Carnegie Selectivity Index classifies institutions into three groups based on the undergraduates admissions rate: 'More selective', 'selective' and 'inclusive'. The three measures of prestige are deeply interrelated, as institutions in the Top 10 of US News & World Report are also more cited and more selective (see Fig. S3). Importantly, in order to capture a wide variety of important institution types we also include mission- and enrollment-driven institutional classifications in our analyses: Historically Black Colleges and Universities (HBCU) and Women's Colleges (WC) (*mission-driven*) and Hispanic Serving Institutions (HSI) (*enrollment-driven*) ) (Kozlowski et al., 2022).. Table S1 provides numbers of papers, number of authors, and number of institutions for each of those groups of universities.

Following our previous work (Kozlowski, Larivière, et al., 2022a) we use topic modeling (Blei et al., 2003) to infer the research topics of articles based on their titles, abstracts, and keywords. We define the topical profile of an intersectional race and gender identity author group as the proportion of papers this author group contributes on each topic with respect to the total number of publications in the topic. Topical profiles can be calculated for both author identity groups and institutional categories. To compare topical profile groups, we use the Spearman rank correlation, as the relation between topical profiles is non-linear. If the correlation between groups (i.e., institutions and intersectional identity) is high, it suggests that they tend to publish on similar topics. We also produce a linear model to predict the effect of authors' identities on scholarly impact (citations and Journal Impact Factor—JIF—). Despite their limitations, citations remain, at the aggregate level, an appropriate indicator for the measurement of the visibility and research impact of papers (Sugimoto & Larivière, 2018), and JIF provides an indication of the selectivity of journals in which they are published (Sugimoto et al., 2013). First, we use an aggregated model to evaluate I) the role of topics, and II) the aggregated effect of institutional prestige:

$$y = \beta_0 + \beta_1 \#authors + \beta_2 career\ age + \sum_i \beta_i\ race\ \&\ gender + \sum_j \beta_j\ institution$$

For each impact indicator (citations and JIF) we build a field-normalized and a topic-normalized dependent variable. The difference between both models reflects on the role that the topical distribution associated with each covariable has on impact. It is worth mentioning that topics cannot be directly coded into the model as covariables because for each field we run a 200-topic model. Topic normalization of the dependent variable allows one to build a single model for all fields. White men are the default race and gender group, while the least prestigious group of each classification is the default for that category. This model has papers as the unit of analysis and is repeated for each prestige classification.

Our second model is built for topic-normalized impact measures; however, instead of adding institutional categories as dummy variables, these groups are split such that a differential model is run for each institutional category separately:

$$y = \beta_{0j} + \beta_{1j} \#authors + \beta_{2j}\, career\ age + \sum_i \beta_{ij}\, race\ \&\ gender$$

By splitting articles into institutional prestige categories and running the linear model for each group, we can examine differential effects on race and gender by institutional prestige and topic.

Given the differences in size among various academic fields, we apply Latent Dirichlet Allocation (LDA) to infer topics within each field separately. For our analysis of author representation (Fig. S4, S6) we use the full dataset. However, for the analysis of correlations between topical profiles (Fig. 1-3), we work at the topic level, and select a specific field for the study. We focus on the LDA model trained on the Social Sciences, Humanities, and Professional fields (and provide alternative results for the field of Health in the SI). We have chosen to highlight these fields as their research topics are deeply intertwined with the social issues that marginalized populations face within and outside of academia. As a result, they offer opportunities for authors to reflect on their socially constructed race and gender identities and the topics they research. That said, our linear models (Fig. 4-5) used for examining scholarly impact uses field- and topic-normalized citations and JIF and takes advantage of the full dataset.

## Results

**Representation and topical profiles of institutions and authors**

Authors with names associated with White men[2] constitute the largest author population across all institution types, with the exception of Women's colleges (WCs), where they are surpassed by names associated with White women (Fig. S4). For all other author groups, the proportion of authors by race and gender are remarkably stable. Relative representation, however, allows for the examination of how certain identities are represented at rates relative to their proportion across all US authors. For Black, Latinx, and women authors, this relative representation demonstrates a strong alignment to mission- and enrollment-driven categorizations of institutions. Specifically, we observe an over-representation of Black men and Black women authors in HBCUs, of Latinx men and Latinx women authors at HSIs, and of women authors in WCs (Fig. S5). Given that HBCUs, HSIs, and WCs are not principally defined by their faculty composition, this finding demonstrates the relationship between institutional mission and author composition, and also serves as a validation of the name-based inference algorithm of race and gender for this level of aggregation. The composition of authors by race and gender also varies as a function of institutional prestige: we observe a relative overrepresentation of Asian authors among institutions with high prestige, and a relative underrepresentation of Black and Latinx authors at prestigious institutions (Fig. S5). The relative overrepresentation of Asian authors should be carefully interpreted given that they are overrepresented in higher education, relative to their representation in the US Census, reinforced by the fact that the majority of Asian doctoral graduates are temporary visa holders (see Fig. S6).

Fig. 1 depicts the correlation between the topical profile of institutions and authors' identities for papers published in the Social Sciences, Humanities, and Professional Fields. The two identity groups—Black women and White men— that present the highest and lowest correlation with the topical profile of Top 10 institutions are presented as examples. The top five topics per institutional group and for each identity were examined and a representative term for each topic is presented in Fig. 1. There is a strong positive correlation between the topical profile of the Top 10 institutions and the topics of White men ($\rho=0.67$), and a strong negative correlation with the topical profile of Black women ($\rho=-0.58$). The correlations are weaker with the Top 100 institutions, where we observe a positive alignment with White men's topical profile ($\rho=0.2$) and a negative relationship for Black women ($\rho=-0.19$). The pattern shifts for lower ranked institutions (Not Top), for which we observe a positive relation with Black women ($\rho=0.48$), and a negative relation with White men ($\rho=-0.53$). Labeling the the most prominent topics of each group helps to illustrate these topic alignments: e.g., Black women publish more on topics related to gender-based violence, literacy, families, and learning. These topics are shared by non-top institutions, who also publish on topics related with tourism, user-perception, and South Korea. On the other hand, both White men and Top 10 institutions share a strong presence with regard to topics related to the market, income-tax, incentive-mechanisms, and the financial sector. Similar results can be seen for Howard University and Harvard University (Fig. S7). While authors at Howard publish more on race, Africa, and African-American culture, all topics in which Black women are also more present; Harvard shares a similar topical space as other Top 10 institutions and White men. These patterns replicate for all women and all men (see Fig. S8). Women from all races have a negative correlation with Top 10 institutions ($-0.58<\rho<-0.42$) and a positive correlation with institutions not in the top ($0.3<\rho<0.48$), while men from all races show a positive correlation with Top 10 institutions ($0.28<\rho<0.67$), and a negative relation with institutions not in the top ($-0.54<\rho<-0.26$). In Health, we observe similar patterns; however, with somewhat lower alignment between institutions and researchers' race and gender (Fig. S9). These results suggest that the topical profile of prestigious institutions is patterned in ways that disproportionately reflect White men's research profiles.

---

[2] From here on 'White men' for simplicity. The same applies for authors with names associated with other race and gender identities.

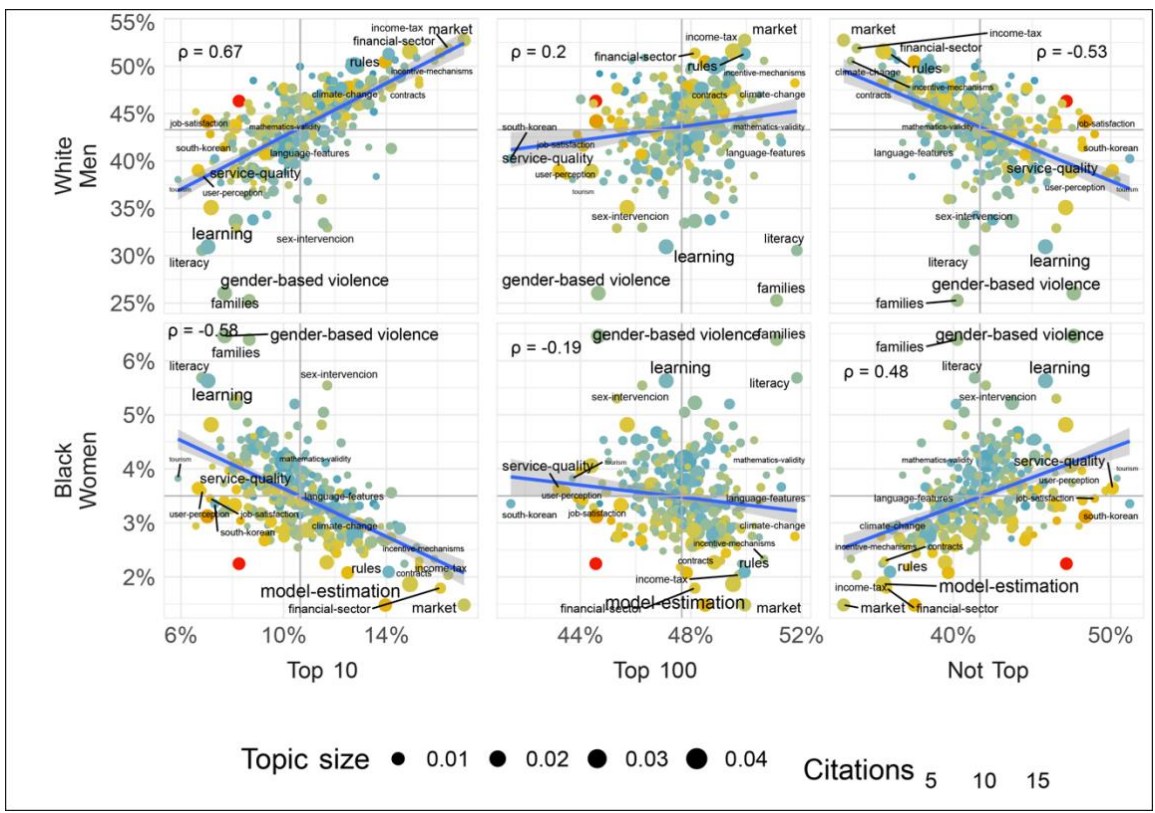

**Figure 1. Topical alignment of institutions and identities.** Proportion of papers in different topics authored by Black Women and White Men (vertical axis) and the percentage of those papers

authored by different institutional groups in the *perceived* prestige groups (horizontal axis). Dot size represents the size of the topic in the corpus associated with the topic, while the dot color represents the average number of citations for that topic. For each subplot, ρ indicates the Spearman correlation, and the blue line is the simple linear regression between the two variables, with the 95% confidence interval. The top 5 topics of each identity and institution group were identified and a representative word is shown.

Fig. 2 provides correlations between all authors' identities and the three levels of *perceived* prestige for Social Sciences, Humanities, and Professional Fields. The topical space occupied by the highest prestige institutions is positively correlated with all men, and negatively correlated with all women. Although the composition of authors by demographic identity varies by institution in relative terms (Fig. S5), the absolute proportion of authors by race and gender remains stable across groups (see Fig. S4), implying that topical profiles are not an artifact of author composition by institution. Moreover, while White men are not more overrepresented in Top 10 institutions than other groups, they still show greater topical alignment with prestigious institutions (ρ=0.67), with a correlation 2.4 times that of Asian men (ρ=0.28) (Fig. S8, and Fig. S9 for Health). Asian women show a negative correlation with Top 10 institutions, even though they are relatively overrepresented in these institutions in terms of authorship. This shows that the relation between identity and topical alignment is not just a projection of underrepresentation, but reflects a more nuanced pattern. Fig. 2 shows how the gender divide is prominent for the topical alignment at prestigious institutions; however, the racial divide creates a spectrum of alignment within gender categories, with White men and Black women at the poles of the intersectional distribution.

These results are robust to other prestige classifications, in that HBCUs, HSIs and Women's colleges are also closely aligned with women's topical profiles (see Fig. S10). As mentioned above, these relations are not an artifact of the composition of authors, and the correlation with respect to the expected topical profile given the demographic composition shows similar results (see Fig. S11). Similarly, these findings hold for other disciplines (see Fig. S12), and considering different levels of disaggregation for institutions (i.e., deciles) (see Fig. S13-14).

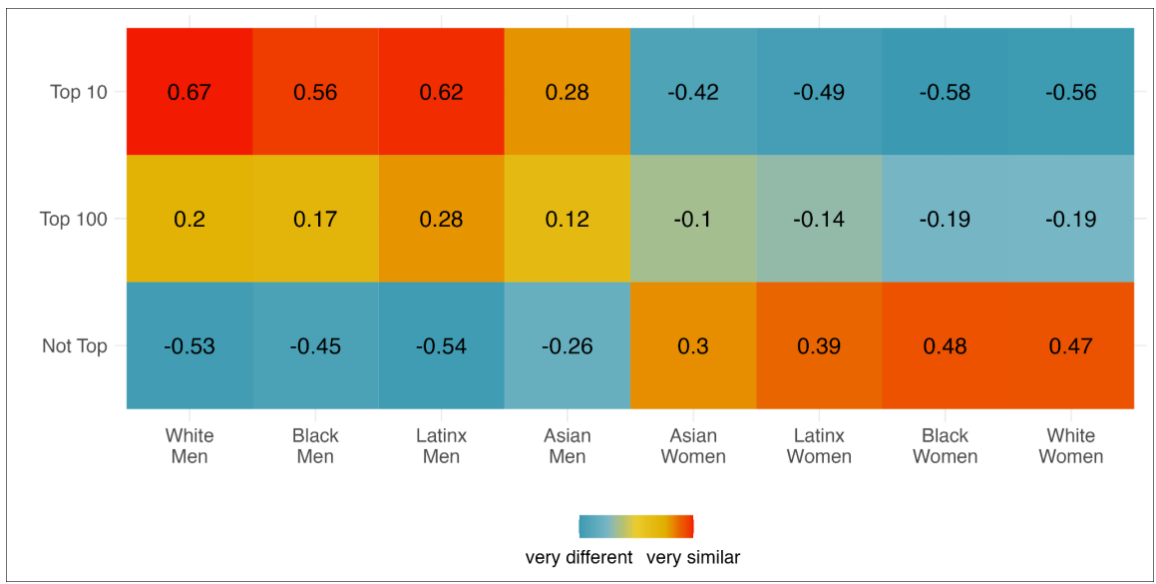

**Figure 2. Spearman correlations between the topic profiles of each author identity and the topical profile of institutional categories for Social Sciences, Humanities and Professional Fields**. Correlations for institutions divided according to *Perceived* prestige from US News & World Report: Top 10 institutions, Top 100 institutions (without the Top 10), and institutions not in the Top 100.

These analyses focused on the correlation between all authors of a given identity and all authors associated with an institutional category. To validate and expand this analysis, we explored the correlation between the topical profile of all authors of an identity group within the institutional category, compared to the topical profile of the identity group overall (i.e., within *race and gender)* (Fig. S15) and compared it to all authors from that institutional category (i.e., within *institution)* (Fig. S16) for Social Sciences, Humanities, and Professional Fields. Important nuances emerge: While for the majority of cases there is a positive relation between the topical profile of authors with respect to their identity across institutions (Fig. S15) (within *race and gender* comparison), which implies that these authors are able to reflect on their identities, there are some exceptions: the topical profile of Asian women at Top 10 institutions is negatively related with the topical profile of Asian women overall, which explains why even though Asian women are overrepresented in prestigious institutions (Fig. S5), their topical profile is negatively aligned with this institutional group (Fig. 2). Likewise, White men at HBCUs have a topical profile that is negatively associated ($\rho$=-0.08) with the general portfolio of White men. All women at Top 10 institutions tend to have profiles that are *less* similar to their general identity portfolio than men at these institutions (Fig. S15). This effect is amplified when one compares each intersectional identity group with that of the population of those institutions (Fig. S16) (*within institution* comparison): e.g., White men at Top 10 institutions are nearly perfectly correlated with the topical profile of their institutions ($\rho$=0.95); whereas Latinx, White, and Black women have a much weaker correlation ($\rho$=0.22, $\rho$=0.23, $\rho$=0.15) respectively. Asian women differ from other women in this regard, demonstrating a stronger topical profile alignment to Top 10 institutions ($\rho$=0.56). The particular alignment of Asian women at prestigious institutions with their institutions rather than their identity might be explained by the migration patterns of authors. While our data does not include precise information on citizenship status, this finding may be explained in part by evidence suggesting that Asian authors are more likely to have migrated to the US for research positions (NSF, 2021a) (see Fig. S6), and given that prestigious institutions host more international scholars (Open Doors, 2022). Similar effects for all identities can be observed in Health (Figs. S17-S18).

Taken together, correlations of topical profiles create a map of the US higher education landscape in which White men and Black Women represent polar ends of a spectrum. Fig. 3 takes these two ends and computes the topical profiles' correlation between them, other intersectional identities, and institutional groups. We identify two particular institutions to illustrate these dimensions: Harvard, a canonical high-prestige institution (which did not have desegregated admissions for women until 1980 (Sugimoto, 2022)), and Howard, a prestigious mission-driven institution founded in 1867 (which opened with five White women students and enrolled the first Black woman in 1884 (Whitford, 2022)).

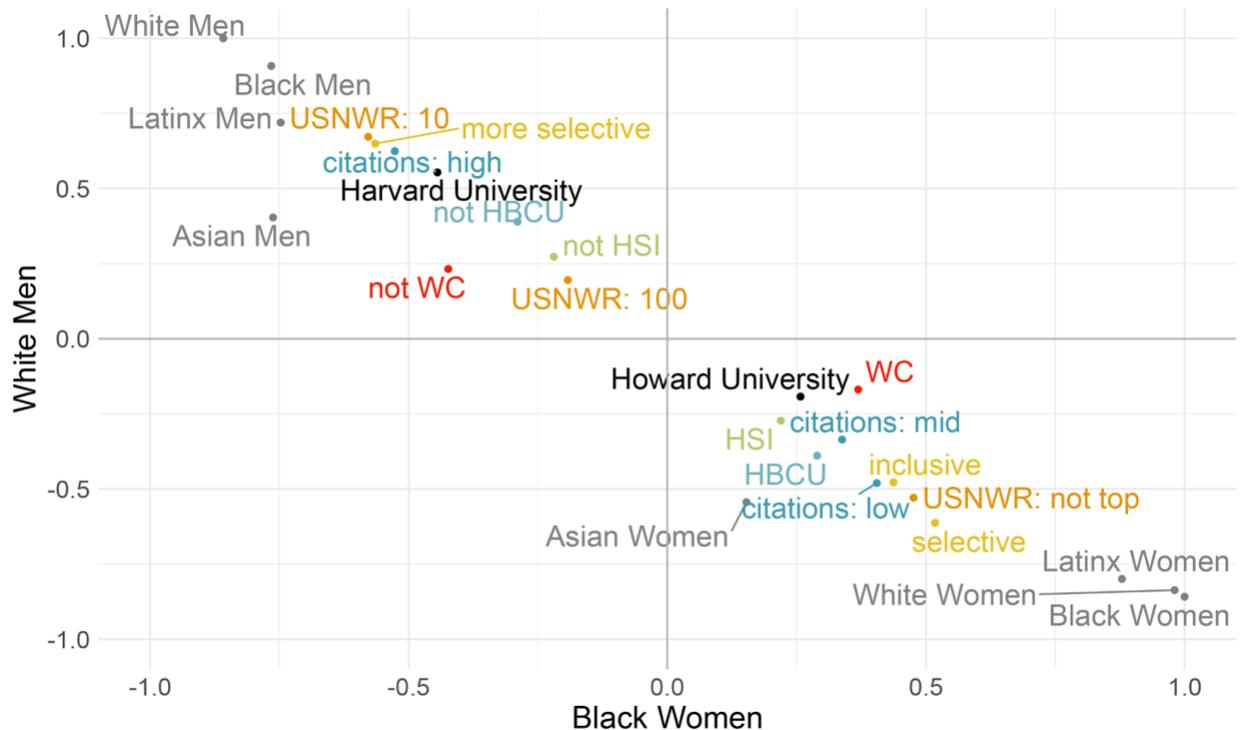

**Figure 3. Scatterplot of Spearman correlations between the topic profiles of each author identity and the topical profile of institutional groups**, for Black Women (horizontal axis) and White men (vertical axis), on Social Science, Humanities, and Professional fields. Institutional groups are: Historically Black Colleges and Universities (HBCU), Hispanic Serving Institutions (HSI), and Women's colleges (WC); US News & World Report: Top 10 institutions (USNWR: 10), Top 100 institutions (without the Top 10) (USNWR: 100), and institutions not in the Top 100 (USNWR: Not Top); Institutions ranked by their average number of citations: Low (0.1, 1.47) (citations: low), Medium (1.48, 1.74) (citations: mid), and High (1.77, 4.07) (citations: high); Institutions according to Selectivity: Carnegie Selectivity Index based on admissions rates (inclusive, selective and more selective).

## Institutions, identities, and impact

The analysis on representation showed a strong racial divide, where White men are overrepresented across institutions (Fig. S4), and Latinx and Black authors are relatively overrepresented in mission- and enrollment-driven institutions and Carnegie's inclusive institutions, and relatively underrepresented in prestigious institutions (Fig. S5). When examining topical profile alignment, we find a strong gender divide, where men are aligned with top institutions and women are aligned with mission- and enrollment-driven institutions and inclusive institutions. In the following section we will study how both race and gender divides relate to authors' impact.

Having established that there is strong alignment between institutional prestige and intersectional author identities in the topical space, we now delve into the question of how these alignments reflect on the impact of research articles. For this, we rely on a set of linear models to show 1) the effect of topical profiles on impact, 2) the role of institutional prestige on impact, and 3) impact gaps by race and gender, and how these are affected by both topics and institutions. With articles as our unit of analysis, two impact indicators are included: citations and JIF. The lowest prestige group and White men are used as reference groups for the institutional and demographic variables. Career age and total number of authors are included as controls.

To measure how the topical profile relates with the articles' impact, we compare the results of field and topic normalizations (Fig. 4). If we restrict the normalization to the field-level, we observe a larger positive effect for institutional prestige and a larger negative effect for marginalized authors (i.e., Black and Latinx men and women, and White women). This means that the topical profile of prestigious institutions contributes to their higher impact, while the topics on which marginalized scholars publish more receive fewer citations and are less published in high impact journals. These results are shared across the different categorizations of prestige, even with prestige as a continuous metric (Fig. S19). Computing the difference in coefficients between each co-variable for the field and topic normalized models provides an indication of the effect of a given topic on the disparities observed (Fig. S20). We observe a positive effect between the topical profiles on scholarly impact for Asian men and women and a penalizing effect for Black and Latinx men and women, and White women. That is, the topics in which Black and Latinx scholars and White women are disproportionately associated are cited at lower rates. Prestigious institutions are also more likely to be associated with topics receiving a higher impact. Therefore, while topical profiles are an important factor for explaining impact gaps between authors and institutions, this gap persists even after normalizing by topics.

The following analysis uses topic-normalized citations and JIF to focus on the relation between identities, institutions, and impact. Our use of topic-normalized impact allows us to leverage from the full dataset of articles published by US first authors. Given that we defined White men as the reference group, the β parameters associated with each identity regarding citations (Fig. 4) can be understood as the proportional *penalty* or *gain* in citations with respect to the topic average for a given intersectional author identity with respect to White men, after controlling for institutional prestige and other covariables (career age, number of co-authors). The model illustrates the strong effect of institutional prestige on impact and provides evidence that author race and gender affect impact, even when controlling for institution type (see Table S2). Specifically, if we examine *perceived* prestige (US news & World report) in Fig. 4; we find that Black and Latinx men ($\beta=-.1$; $\beta=-.09$, respectively) and women ($\beta=-.17$; $\beta=-.13$, respectively) receive fewer citations, on average, and publish in journals with lower JIFs than White men (Black women: $\beta=-.1$; Black men: $\beta=-.07$; Latinx women $\beta=-.05$; and Latinx men: $\beta=-.02$). Notably, the negative effect of author race and gender for both citations and JIF is most pronounced for Black and Latinx women. Asian men and women publish in journals with higher JIFs than White men ($\beta=.06$; $\beta=.03$, respectively), and Asian men receive higher average citations ($\beta=.09$). White, Latinx, and Black women receive fewer citations ($\beta=-.07$; $\beta=-.13$; $\beta=-.17$; respectively) and publish in journals with lower JIFs than White men ($\beta=-.03$; $\beta=-.05$; $\beta=-.1$; respectively).

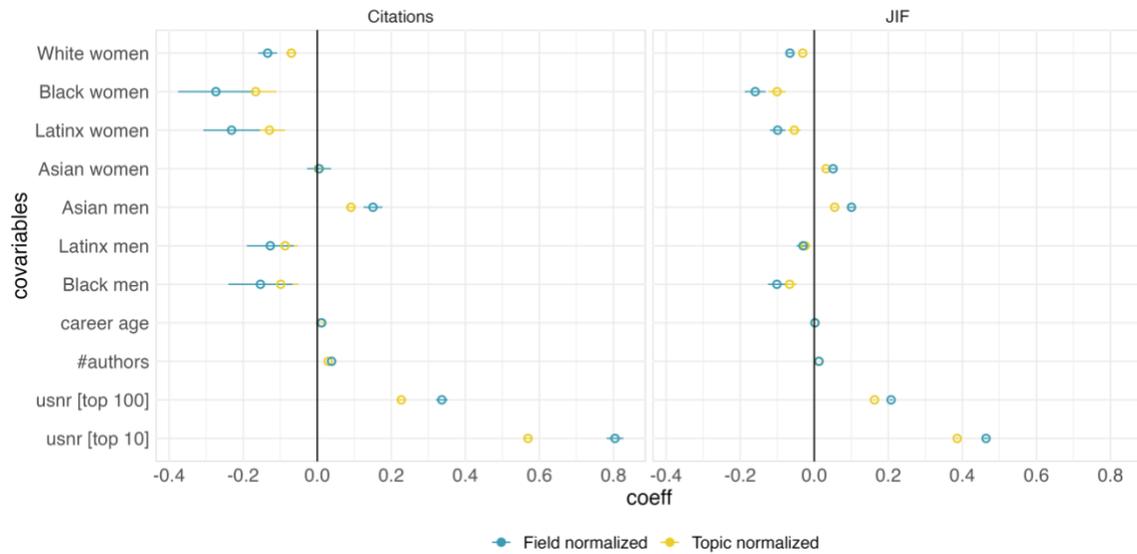

**Figure 4. Parameters of aggregated linear regression models for citations and JIF.** The reference group for our intersectional race by gender identity variables is White men, with the number of co-authors and career age serving as controls. Results for the *perceived* (US News & World Report): Top 10 institutions, Top 100 institutions (without the Top 10) and institutions not in the Top 100 (default group). Values in blue represent the field-normalized models, while values in yellow have topic-normalized dependent variables. A topic model with 200 topics was built for each field to perform the normalization. Error bars represent the 95% confidence interval.

Impact gaps by race and gender remain after controlling for institutional prestige and topics. However, this does not necessarily imply that the impact gap is equal across institutions. We aim to understand not only how institutional prestige boosts impact, but also how it relates to impact gaps. To examine this phenomena, we split the population of papers given their institutional affiliation into our three rank-ordered prestige categories (see Fig. S21 and Table S3 for alternative prestige metrics) and run the topic-normalized citations and JIF linear models for each group. As shown in Fig. 5, author race and gender have a larger effect on scholarly impact at the most prestigious institutions; specifically, for Latinx, Black, and White women. These identities have lower citations, on average, and publish in journals with lower JIF at all institutions, but the effect is most pronounced at institutions with higher *perceived* prestige.

For White women, the penalty on citations is 6.6 times larger at high prestige institutions than in low prestige institutions ($\beta=-.2$; $\beta=-.03$ respectively), while for Latinx women is 6 times higher ($\beta=-.36$; $\beta=-.06$ respectively). Although the large variability makes results for Black women in the top 10 institutions non-significant, we can still see the same patterns between top 100 and non-top institutions, where the negative impacts are larger than for any other demographic group ($\beta=-.22$; $\beta=-.11$, respectively). Similar patterns are also observed for the JIF, where Black authors are penalized most at top institutions (women: $\beta=-.19$; men: $\beta=-.14$) and where those penalties are considerably larger than in non-top institutions (women: $\beta=-.09$; men: $\beta=-.06$). Notably, White women publish in journals with an almost equal topic-normalized JIF as their White men peers in non-top institutions, but not in top institutions ($\beta=-.01$; $\beta=-.11$ respectively).

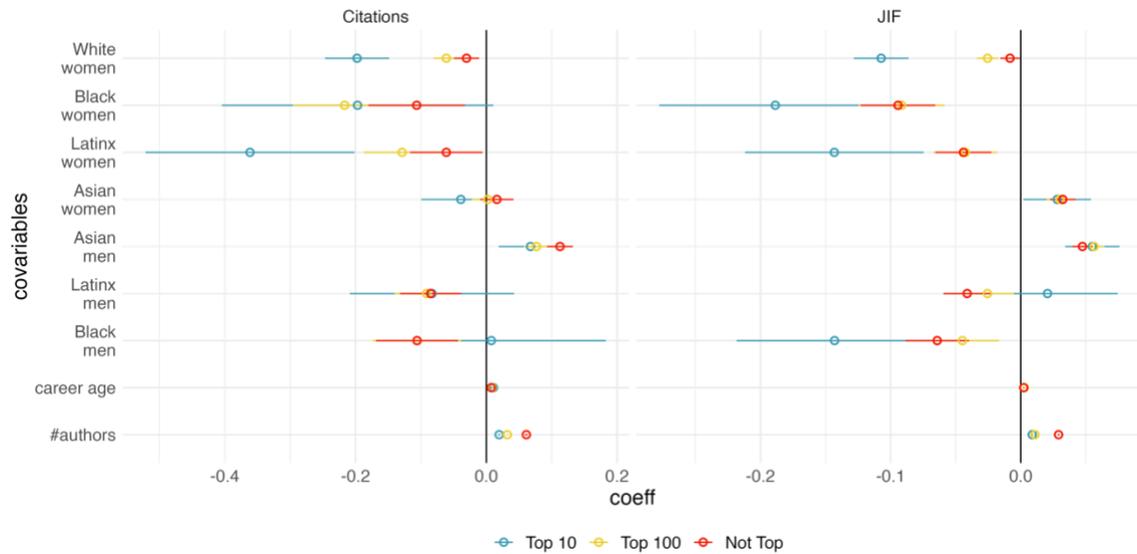

**Figure 5. Parameters of linear regression models predicting the topic and year normalized citations and JIF, for subsets of institutions**. The reference group for our intersectional race and gender identity variables is White men, with the number of co-authors and career age serving as controls. Results for the three linear models run on the data grouped by *perceived* prestige (US News & World Report): Top 10 institutions, Top 100 institutions (without the Top 10) and institutions not in the Top 100.

To analyze the intersectional disparities in scientific impact at prestigious institutions, we compare, for each identity, the difference in citations and JIFs in high and low prestige institutions, with respect to the middle group (Fig. S22). Overall, men authors experience greater positive and negative effects based on their institutional affiliation than women authors across each prestige indicator (i.e., *perceived*, *research,* and *selective*) for both citations and JIF. If we examine *perceived* prestige for example, relative to Top 100 institutions; Black men experience a 42.4% citation gain, and 23.3% JIF gain at Top 10 institutions; while Black men at Not Top institutions experience a 25.8% citation penalty, and 17.7% JIF penalty. Similar patterns emerge for women, but the gains and penalties are not as substantial. Latinx women experience a 21.7% citation gain, and 17.5% JIF gain at Top 10 institutions; and a 20.5% citation penalty, and 16.1% JIF penalty at Not Top institutions.

Citation gaps between White men and other intersectional identities demonstrate a trend towards increased marginalization: increasing institutional prestige is associated with increasing inequalities in citations (Table S4) and JIFs (Table S5). For example, Latinx women have a difference of 12 percentage points in the citation gap between Not Top institutions (-7.7% citation penalty), and Top 10 institutions (-19.84% citation penalty) (Table S4). The difference is smaller for White women (-6.8% and -15.1% in Not Top and Top 10 institutions respectively) and Black women (-9% and -15.5% in Not Top and Top 10 institutions respectively). This disparity holds for JIF, where we see in Not Top institutions a gap of -3.8%, -3%, -1.9% for Black, Latinx and White women, respectively, and of -9.2%, -8.8% and -8.1% on Top 10 institutions for those same groups (Table S5). The results imply that for Black, Latinx and White women, the differences in citations and JIF between institutions are not as substantial between institutional groups as they are for White men (Fig. S22). Similar patterns emerge for field-normalized citations (Tables S6-S7). This model provides clear evidence that being affiliated with a prestigious university has a positive impact on the citations and JIFs of all authors; however, this advantage is largest for White men.

# Discussion

Institutions of higher education are increasingly being scrutinized for their role in reproducing inequalities in science (Clauset et al., 2015; Wapman et al., 2022). Economic and symbolic capital are highly concentrated in these institutions (Sugimoto, 2022; Whitford, 2022): few institutions control the production of faculty (Wapman et al., 2022), with research from these institutions having outsized scientific impact (Way et al., 2019). Policy interventions at prestigious institutions, therefore, have the opportunity to significantly alter the scientific landscape. This study provides an intersectional analysis of the relationship between the prestige of institutions and scientific impact. We provide evidence that elite institutions have a lower representation of Black and Latinx authors, have a higher topical alignment with White men, and reproduce the largest impact gaps. Of course, race and gender identities intersect with other dimensions of inequality, especially class. Our data limitations inhibit us to include this dimension of analysis, but it has been shown that socioeconomic roots of authors affect their career development (Morgan et al., 2022), and Black and Latinx populations are disproportionately associated with lower socioeconomic status in the US (Oliver & Shapiro, 2006).

In our previous work, we found that the topical profile of non-white authors have a greater focus on issues of direct relevance to their racialized and gendered identities —e.g., racial discrimination, migration, and gender-based violence (Kozlowski, Larivière, et al., 2022a). In the present work, we find that HBCUs, HSIs and WCs have research profiles that are closely related to the topical profiles of Black, Latinx, and White women. However, as we climb the ladder of institutional prestige[3], we observe a sharp decline in the relative presence of these topics. Far from being a consequence of the composition of their respective authorship, this patterns reflects a more complex phenomenon: not only do authors from historically minoritized groups—particularly those at prestigious institutions—have a topical profile that differs from the dominant topical profile at their home institutions (i.e., within *institutional* difference), they also have a topical profile that differs from the dominant topical profile of other authors within their own intersectional identities across institutions (i.e., within *race and gender* difference).

Disparities in citation by topic are particularly disadvantageous for Black and Latinx men and women and White women (Fig. 4). However, even when controlling for topic and institutional prestige (Fig. 4-5), these populations receive fewer citations and publish in journals of lower JIF. This suggests that topic selection alone does not fully explain disparities in citations. We corroborate the known relationship between impact and prestige (Hagstrom, 1971) and note that authors from all identities affiliated with prestigious institutions receive an impact advantage. However, these advantages are not distributed equally. It is at the institutions of highest prestige that we observe the largest disparities in impact, particularly for Black, Latinx, and White women. This suggests that, even when controlling for topic choice and institutional placement, there remains a disparity in impact for Black, Latinx, and White women. One possible explanation for the smaller impact gap between men and women authors at less prestigious institutions is the relative under-placement of women in faculty positions (Clauset et al., 2015). Similar employment mechanisms could be driving other trends observed, such as the topical misalignment between Asian men and women in HBCUs (Betsey, 2007). Another possible explanation is that, given the skewness of impact distribution, differences among high impact authors in prestigious institutions tend to be nominally higher. Regardless of the explanation, the disproportionate advantage for White men at prestigious institutions further codifies stratification at the intersection of institutional prestige and authors' identity.

---

[3] Three HSIs rank within the top 100 of US News & World Report 2021 rankings (University of California, Santa Barbara; University of California, Riverside; and Texas A&M University-College Station); just one HBCU meets this threshold (Howard University); and there are no Women's Colleges within this list.

To promote topical diversity in science, we need strategic shifts at institutional and federal levels. Federal agencies are the largest supporters of academic R&D in the US (i.e., 53%, ~ $45 billion); however, more than 20% of research funding is also derived from within academic institutions themselves (~ $21 billion) (NCSES, 2019). There is tremendous variation in the degree to which academic R&D is institutionally and federally supported. For example, Howard University, the only HBCU to be ranked in the top 100 of US News & World Report, had nearly $45 million in R&D expenditures in 2021, of which 65% came from federal funding sources (~$30 million) and 24% from internal sources (~$11 million) (NCSES, 2020a; USNWR, 2021). By comparison, Harvard University, ranked second in US News & World Report, had approximately $1.2 billion in R&D expenditures in 2021, of which 49% (~$601 million) originated from federal sources, and 32% (~$390 million) from internal sources (NCSES, 2020b; USNWR, 2021). In relative terms, Harvard is less reliant on federal funding for research than Howard University, suggesting that the institution has greater ability to strategically organize funding towards marginalized topics and to support the work of minoritized scholars (through funding, hiring, promotion, amplification, and mentorship policies). Institutions with higher reliance on federal funding should advocate for change within these agencies; acknowledging systemic disparities in funding (Chen et al., 2022; Hoppe et al., 2019) and recommending new practices for more equitable evaluation (Hunt et al., 2022).

Editors, journals, and publishers are also pivotal actors in this space. There is a nontrivial and reinforcing relationship between funding and publishing (Győrffy et al., 2020; Jacob & Lefgren, 2011). Therefore, journals' broader acceptance of topics of salience to marginalized communities is likely to have effects on both who and what is funded. Editors can ensure that they are reflexive in considering the ways in which they may promote sexist or racist discourse and imagery in their coverage of work (Nature, 2022) and work to mitigate bias through the selection of more diverse teams of reviewers (Murray et al., 2019). These actions may also have a cascading effect in promoting other aspects of reflective and robust scientific practices that serve to elevate the work of minoritized scholars (Dworkin et al., 2020; Kwon, 2022).

The Matthew-Matilda effect refers to cumulative advantages (Merton, 1968) and disadvantages (Rossiter, 1993) in science. At the institutional level, we observe a Howard-Harvard effect, in which mission-driven institutions not only show a higher proportion of students, faculty and authorship from the communities they serve, but also a topical profile that fills the knowledge gap produced by intersectional inequalities in science (Kozlowski, Larivière, et al., 2022a). Prestigious institutions, on the other hand, present both an underrepresentation of Black and Latinx authors, a men-aligned topical profile, and the largest impact gaps. The US higher education system, and the actors that support it, are called to reduce the systemic marginalization of particular identities and topics of greatest salience for these populations.

# References and Notes

Bertolero, M. A., Dworkin, J. D., David, S. U., Lloreda, C. L., Srivastava, P., Stiso, J., Zhou, D., Dzirasa, K., Fair, D. A., Kaczkurkin, A. N., Marlin, B. J., Shohamy, D., Uddin, L. Q., Zurn, P., & Bassett, D. S. (2020). *Racial and ethnic imbalance in neuroscience reference lists and intersections with gender* (p. 2020.10.12.336230). bioRxiv. https://doi.org/10.1101/2020.10.12.336230

Betsey, C. L. (2007). Faculty Research Productivity: Institutional and Personal Determinants of Faculty Publications. *The Review of Black Political Economy*, *34*(1–2), 53–85. https://doi.org/10.1007/s12114-007-9004-9

Blei, D. M., Ng, A. Y., & Jordan, M. I. (2003). Latent dirichlet allocation. *The Journal of Machine Learning Research*, *3*(null), 993–1022.

Chen, C. Y., Kahanamoku, S. S., Tripati, A., Alegado, R. A., Morris, V. R., Andrade, K., & Hosbey, J. (2022). *Decades of systemic racial disparities in funding rates at the National Science Foundation*. OSF Preprints. https://doi.org/10.31219/osf.io/xb57u

Clauset, A., Arbesman, S., & Larremore, D. B. (2015). Systematic inequality and hierarchy in faculty hiring networks. *Science Advances*, *1*(1), e1400005. https://doi.org/10.1126/sciadv.1400005

Crenshaw, K. (1991). Mapping the Margins: Intersectionality, Identity Politics, and Violence against Women of Color. *Stanford Law Review*, *43*(6), 1241–1299. https://doi.org/10.2307/1229039

Dworkin, J., Zurn, P., & Bassett, D. S. (2020). (In)citing Action to Realize an Equitable Future. *Neuron*, *106*(6), 890–894. https://doi.org/10.1016/j.neuron.2020.05.011

Erosheva, E. A., Grant, S., Chen, M.-C., Lindner, M. D., Nakamura, R. K., & Lee, C. J. (2020). NIH peer review: Criterion scores completely account for racial disparities in overall impact scores. *Science Advances*, *6*(23), eaaz4868. https://doi.org/10.1126/sciadv.aaz4868

Ginther, D. K., Schaffer, W. T., Schnell, J., Masimore, B., Liu, F., Haak, L. L., & Kington, R. (2011). Race, Ethnicity, and NIH Research Awards. *Science*, *333*(6045), 1015–1019. https://doi.org/10.1126/science.1196783

Győrffy, B., Herman, P., & Szabó, I. (2020). Research funding: Past performance is a stronger predictor of future scientific output than reviewer scores. *Journal of Informetrics*, *14*(3), 101050. https://doi.org/10.1016/j.joi.2020.101050

Hagstrom, W. O. (1971). Inputs, Outputs, and the Prestige of University Science Departments. *Sociology of Education*, *44*(4), 375. https://doi.org/10.2307/2112029

Hoppe, T. A., Litovitz, A., Willis, K. A., Meseroll, R. A., Perkins, M. J., Hutchins, B. I., Davis, A. F., Lauer, M. S., Valantine, H. A., Anderson, J. M., & Santangelo, G. M. (2019). Topic choice contributes to the lower rate of NIH awards to African-American/black scientists. *Science Advances*, *5*(10), eaaw7238. https://doi.org/10.1126/sciadv.aaw7238

Hunt, L., Nielsen, M. W., & Schiebinger, L. (2022). A framework for sex, gender, and diversity analysis in research. *Science*, *377*(6614), 1492–1495. https://doi.org/10.1126/science.abp9775

Jacob, B. A., & Lefgren, L. (2011). The impact of research grant funding on scientific productivity. *Journal of Public Economics*, *95*(9), 1168–1177. https://doi.org/10.1016/j.jpubeco.2011.05.005

Kozlowski, D., Doshi, S., Rangwala, A., Sugimoto, C. R., Larivière, V., & Monroe-White, T. (2022, September 7). *Applying an Intersectional Lens to Author Composition at Women's Colleges, Historically Black Colleges and Universities, and Hispanic Serving Institutions in the United States*. STI, Granada. https://orbilu.uni.lu/handle/10993/52219

Kozlowski, D., Larivière, V., Sugimoto, C. R., & Monroe-White, T. (2022a). Intersectional inequalities in science. *Proceedings of the National Academy of Sciences*, *119*(2), e2113067119. https://doi.org/10.1073/pnas.2113067119

Kozlowski, D., Larivière, V., Sugimoto, C. R., & Monroe-White, T. (2022b, October 9). Race and gender homophily in collaborations and citations. *Metrics 2022: ASIS&T Virtual Workshop on Informetrics and Scientometrics Research*. Metrics 2022. https://orbilu.uni.lu/handle/10993/52536

Kozlowski, D., Murray, D. S., Bell, A., Hulsey, W., Larivière, V., Monroe-White, T., & Sugimoto, C. R. (2022). Avoiding bias when inferring race using name-based approaches. *PLOS ONE*, *17*(3), e0264270. https://doi.org/10.1371/journal.pone.0264270

Kwon, D. (2022). The rise of citational justice: How scholars are making references fairer. *Nature*, *603*(7902), 568–571. https://doi.org/10.1038/d41586-022-00793-1

LaBerge, N., Wapman, K. H., Morgan, A. C., Zhang, S., Larremore, D. B., & Clauset, A. (2022). *Subfield prestige and gender inequality in computing* (arXiv:2201.00254). arXiv. http://arxiv.org/abs/2201.00254

Langin, K. (2020). LGBTQ researchers say they want to be counted. *Science*, *370*(6523), 1391–1391. https://doi.org/10.1126/science.370.6523.1391

Larivière, V., Desrochers, N., Macaluso, B., Mongeon, P., Paul-Hus, A., & Sugimoto, C. R. (2016). Contributorship and division of labor in knowledge production. *Social Studies of Science*, *46*(3), 417–435. https://doi.org/10.1177/0306312716650046

Larivière, V., Ni, C., Gingras, Y., Cronin, B., & Sugimoto, C. R. (2013). Bibliometrics: Global gender disparities in science. *Nature*, *504*(7479), Article 7479. https://doi.org/10.1038/504211a

McGee, E. O., Parker, L., Taylor, O. L., Mack, K., & Kanipes, M. (2021). HBCU Presidents and their Racially Conscious Approaches to Diversifying STEM. *Journal of Negro Education*, *90*(3), 288–305.

Merton, R. K. (1968). The Matthew Effect in Science. *Science*, *159*(3810), 56–63. https://doi.org/10.1126/science.159.3810.56

Morgan, A. C., Economou, D. J., Way, S. F., & Clauset, A. (2018). Prestige drives epistemic inequality in the diffusion of scientific ideas. *EPJ Data Science*, *7*(1), Article 1. https://doi.org/10.1140/epjds/s13688-018-0166-4

Morgan, A. C., LaBerge, N., Larremore, D. B., Galesic, M., Brand, J. E., & Clauset, A. (2022). Socioeconomic roots of academic faculty. *Nature Human Behaviour*, 1–9. https://doi.org/10.1038/s41562-022-01425-4

Murray, D., Siler, K., Larivière, V., Chan, W. M., Collings, A. M., Raymond, J., & Sugimoto, C. R. (2019). *Author-Reviewer Homophily in Peer Review* (p. 400515). bioRxiv. https://doi.org/10.1101/400515

Nature. (2022). How Nature contributed to science's discriminatory legacy. *Nature*, *609*(7929), 875–876. https://doi.org/10.1038/d41586-022-03035-6

NCSES, N. C. for S. and E. S. (2020a). *Harvard—Source: National Center for Science and Engineering Statistics, Higher Education R&D Survey. Total R&D expenditures, by source of funds and R&D field: 2020*. https://ncsesdata.nsf.gov/profiles/

NCSES, N. C. for S. and E. S. (2020b). *HOWARD -- Source: National Center for Science and Engineering Statistics, Higher Education R&D Survey. Total R&D expenditures, by source of funds and R&D field: 2020*. https://ncsesdata.nsf.gov/profiles/

NCSES, N. C. for S. and E. S. (2019). *Higher Education Research and Development Survey (HERD). Science and Engineering Indicators. Academic R&D expenditures, by source of support: FY 2019 NSB-2021-3*. https://ncses.nsf.gov/pubs/nsb20213/financial-resources-for-academic-r-d

NSF, N. S. F. (2021a). *Doctorate Recipients from U.S. Universities: 2019*. https://ncses.nsf.gov/pubs/nsf21308/table/19

NSF, N. S. F. (2021b). *Women, Minorities, and Persons with Disabilities in Science and Engineering*. https://ncses.nsf.gov/pubs/nsf21321/report/executive-summary

Oliver, M., & Shapiro, T. M. (Eds.). (2006). *Black Wealth / White Wealth: A New Perspective on Racial Inequality, 2nd Edition* (2nd edition). Routledge.

Open Doors. (2022). *IIE Open Doors / Institutions Hosting the Most Scholars*. IIE Open Doors / Institutions Hosting the Most Scholars. https://opendoorsdata.org/data/international-scholars/institutions-hosting-the-most-scholars/

Owens, E. W., Shelton, A. J., Bloom, C. M., & Cavil, J. K. (2012). The Significance of HBCUs to the Production of STEM Graduates: Answering the Call. *Educational Foundations*, *26*, 33–47.

Peng, H., Teplitskiy, M., & Jurgens, D. (2022). *Author Mentions in Science News Reveal Widespread Disparities Across Name-inferred Ethnicities* (arXiv:2009.01896). arXiv. https://doi.org/10.48550/arXiv.2009.01896

Ross, E. (2017). Gender bias distorts peer review across fields. *Nature*. https://doi.org/10.1038/nature.2017.21685

Ross, M. B., Glennon, B. M., Murciano-Goroff, R., Berkes, E. G., Weinberg, B. A., & Lane, J. I. (2022). Women are credited less in science than men. *Nature*, *608*(7921), Article 7921. https://doi.org/10.1038/s41586-022-04966-w

Rossiter, M. W. (1993). The Matthew Matilda Effect in Science. *Social Studies of Science*, *23*(2), 325–341.

Sax, L. J., Berdan Lozano, J., & Korgan, C. (2014). *Who Teaches at Women's Colleges?*

Sugimoto, C. R. (2022). Narrow hiring practices at US universities revealed. *Nature*, *610*(7930), 37–38. https://doi.org/10.1038/d41586-022-03065-0

Sugimoto, C. R., & Larivière, V. (2018). *Measuring Research: What Everyone Needs to Know* (1st edition). Oxford University Press.

Sugimoto, C. R., Larivière, V., Ni, C., & Cronin, B. (2013). Journal acceptance rates: A cross-disciplinary analysis of variability and relationships with journal measures. *Journal of Informetrics*, *7*(4), 897–906. https://doi.org/10.1016/j.joi.2013.08.007

Teich, E. G., Kim, J. Z., Lynn, C. W., Simon, S. C., Klishin, A. A., Szymula, K. P., Srivastava, P., Bassett, L. C., Zurn, P., Dworkin, J. D., & Bassett, D. S. (2022). Citation inequity and gendered citation practices in contemporary physics. *Nature Physics*, *18*(10), Article 10. https://doi.org/10.1038/s41567-022-01770-1

USBC, U. C. B. (2016). *Frequently Occurring Surnames from the 2010 Census*. Census.Gov. https://www.census.gov/topics/population/genealogy/data/2010_surnames.html

USNWR. (2021). *US News and World Report ranking*.

Wapman, K. H., Zhang, S., Clauset, A., & Larremore, D. B. (2022). Quantifying hierarchy and dynamics in US faculty hiring and retention. *Nature*, *610*(7930), Article 7930. https://doi.org/10.1038/s41586-022-05222-x

Way, S. F., Morgan, A. C., Larremore, D. B., & Clauset, A. (2019). Productivity, prominence, and the effects of academic environment. *Proceedings of the National Academy of Sciences*, *116*(22), 10729–10733. https://doi.org/10.1073/pnas.1817431116

Whitford, E. (2022, February 18). *College Endowments Boomed in Fiscal 2021*. Inside Higher Ed. https://www.insidehighered.com/news/2022/02/18/college-endowments-boomed-fiscal-year-2021-study-shows

Zhang, S., Wapman, K. H., Larremore, D. B., & Clauset, A. (2022). Labor advantages drive the greater productivity of faculty at elite universities. *Science Advances*, *8*(46), eabq7056. https://doi.org/10.1126/sciadv.abq7056

Zuberi, T. (2000). Deracializing Social Statistics: Problems in the Quantification of Race. *The ANNALS of the American Academy of Political and Social Science*, *568*(1), 172–185. https://doi.org/10.1177/000271620056800113

**Supplementary Information for**

**The Howard-Harvard effect: Institutional reproduction of intersectional inequalities**

**Authors:** Diego Kozlowski, Thema Monroe-White, Vincent Larivière, Cassidy R. Sugimoto*

*Corresponding author.

**Email:** sugimoto@gatech.edu

**Supplementary Information Text**

**Materials and Methods**

Data

Our dataset consists of 5,431,451 articles published between 2008 and 2020 and indexed in the Web of Science (WOS), for which the first author carries a U.S. affiliation, and the distinct 4,713,444 first authors affiliated with these articles. These articles were associated with 261,336 distinct institution name strings, which were cleaned to assign papers to specific universities. The cleaning process for institutions consisted of two tasks: first, normalizing the multiple strings by which the name of the same university appears in WOS; second, building a crosswalk between institutions' names as they appear in WOS and in the Carnegie list of institutions. Both tasks were first conducted algorithmically, and then checked manually. The institutions selected for the manual cleaning followed a double criterion: first, we considered all institutions names in WOS that appeared 500 times or more. Given that this work also focuses on HBCUs, HSIs, and Women's Colleges, we did a second round of manual cleaning for names in WOS that partially matched those of the institutions in Carnegie from these groups, with a smaller threshold of 25 instances. This latter step allows us to triple our coverage of these institutions. After cleaning, the final dataset consists of 4,553,335 articles, 3,441,264 U.S. first authors, and 685 universities, which covers 84% of articles and 73% of authors contained in the original dataset. Out of the 685 colleges and universities analyzed, 62 are HBCUs (out of 100 in Carnegie), 127 are HSIs (out of 803), and 25 are WC (out of 34). The lack of coverage of all institutions may in part be due to a low signal in WoS for many HBCUs, WCs, and HSIs. In addition, we took a manual approach to retrieving all articles with Tribal Colleges, which are also mission-driven institutions categorized by the Carnegie classification. However, the low volume of articles retrieved (500) for those institutions—which is a finding in itself—did not allow us to perform further analyses. This is an acknowledged limitation of the present work. It is important to also note that within these institutional categories, the imbalance in the number of publications across institutions means that the results are driven by the leading institutions of each category. For example, the Top 10 most productive HBCUs published 70% of the articles of the group, while the 40 least productive accounts for less than 9% of the articles. For HSIs, the Top 10 institutions account for 73% of the papers, while the remaining 117 account for only 27%. In Women's colleges, the Top 10 institutions published 90% of the articles, while the remaining 18 published 10%.

Institutional prestige is a key variable of this analysis. We rely on three different indicators of institutional prestige: US News & World Report ranking, the historical average of field-normalized numbers of citations of institutions, and Carnegie's selectivity index. For each of these, we split the institutions into three groups: high, middle and low prestige.

US News & World report use a compound of factors such as graduation rates, faculty resources, and undergraduate academic reputation to determine the ranking of the—in their terms—best colleges in US[4]. We used the 2022 edition of the report and search their website[5] to match the Top 100 institutions with our curated WOS database. With this information we split the universities between those in the Top 10 of the ranking (11 universities, given ties), those between the Top 10 and Top 100 (89 universities) and those that fall outside the Top 100 (584 universities). Research production remains uneven within this group, with Top 10 institutions accounting for 17% of articles, and the Top 100 accounting for 47% of articles. Given the widespread use of this ranking by society to form expectations about institutions, we consider this to be a classification of *perceived prestige*.

We also used the historical average of field-normalized number of citations (Waltman & van Eck, 2019) by institution. For this, we use all WOS-indexed articles published by universities between 1980 and 2019, and the field- and year-normalized citations. To build the high/medium/low average citations groups we used a weighted version of quantiles that considers the number of publications, in order to build groups of similar size. Highly cited institutions (60 universities) have between 1.77 and 4.07 normalized citations per article. Medium cited institutions (78 universities) move between 1.48 and 1.74 normalized citations, while low cited institutions (547) have between 0.1 and 1.47 citations on average. Each of the three groups account for roughly 33% of articles each (see Table S1). This citation-based classification of prestige can be labeled as *research prestige*, as it is based on the research impact of papers from each university. We also build alternative classifications of impact, using the total number of citations, the proportion of paper an institution has in the top 1%, 5% and 10% most cited articles. All of these classifications yield similar results, and hence we decided to use the historical average number of normalized citations for simplicity.

As a third approach to the prestige of institutions, we used the Carnegie Selectivity index (Carnegie, 2022), a metric built by the Carnegie Classification of Institutions of Higher Education which can be retrieved on the official website[6], which divides universities according to their undergraduate admission rates. Those are divided as "inclusive" (196 universities), "selective" (206 universities)

---

[4] https://www.usnews.com/education/best-colleges/articles/how-us-news-calculated-the-rankings (retrieved 29/08/2022)

[5] https://www.usnews.com/best-colleges/rankings/national-universities (retrieved 29/08/2022)

[6] http://carnegieclassifications.acenet.edu/downloads/CCIHE2021-PublicDataFile.xlsx. Version 9, Accessed November 19, 2022)

and "more selective" (187 universities). This perspective focuses on the elitism of the institution within the student population, and we call it *selectivity prestige*. We did not use Carnegies' Basic Classification because R1 institutions account for a great majority of research papers, generating an imbalanced dataset that is unable to show differences within the R1 universities.

Each of these three operationalisations gives a partial view of prestige. US News & World report is a widely regarded ranking by the US society overall. It therefore affects the perception that the broader society has about the prestige of an institution. The historical average number of citations shows the impact that an institution has within the scientific community. The selectivity index shows the elitism of the student population. The similarity of the outcomes on these three levels gives robustness to the analysis. The size of the groups differs across classifications (see table S1), which has an impact on the behavior of the middle groups, as these depend on the thresholds that define them (see for example Fig. S12). The US News & World Ranking shows a narrower definition of top institution, including only 11 institutions and less than 1M papers, while the top group based on selectivity gathers more than 3M papers and is the biggest of the three groups. Conversely, the bottom group based on selectivity gathers 262,617 articles, while institutions not in the top 100 of US News & World report gather 1.9M articles. Given the nature of the indicator, the three groups based on the average number of citations retrieve between 1.75M and 1.83M articles. These thresholds are arbitrary cuts of the prestige dimension and are simply heuristics for our analysis. Given that the construction of the selectivity prestige and perceived prestige are taken from the respective sources, we decided to build the impact prestige with a roughly equal number of publications per bin. In this way, the perceived prestige shows the most imbalanced groups, the selectivity prestige is balanced in number of universities but imbalanced in number of articles (given the productivity difference), and the impact prestige is imbalanced in the number of universities but balanced in number of articles (see table S1 and Fig. S3). We also build a decile version of the impact prestige (see Fig. S13-14) and worked for the linear model on the continuous version of this variable. The results show similar behaviors on the high and low prestige groups across categories, allowing a robust interpretation of the results.

Following Kozlowski et al. (Kozlowski, Murray, et al., 2022), authors of the selected papers were assigned a race based on the association between their family names and race found in the US census data (USBC, 2016).We avoid using thresholds to assign authors to a single racial group as this approach significantly underestimates the number of Black and Latinx authors. The root cause of this underrepresentation is the prevalence of common family names between the White and Black populations in the US, which is a legacy of slavery (Furstenberg, 2007). Additionally, we refrain from using given names for racial inference as the US census does not provide such information, and alternative data sources, such as mortgage applications (Tzioumis, 2018), have been shown to under-represent Black and Latinx populations. For names that are not present in the census data, we impute the mean of the distribution of names in WOS that do appear on the census. This is because using the census average would assume that the census data and the WOS populations are equal, which is not the case. However, using census data to infer names still assumes that the distributions of authors and the US population are equivalent. As Black and Latinx authors are underrepresented in WOS with respect to the census, our approach may tend to overestimate the proportion of Black and Latinx authors, as was shown in the previous literature (Laberge et al., 2022). Thus, the results presented here are likely to be a lower bound estimate. The potential issues with our method are twofold. Firstly, in terms of representation, our approach may overestimate the actual population of Black and Latinx scholars. Secondly, due to the overlap of family names between Black and White authors, we expect wider actual differences in the distribution of authors across institutions and topics between these two populations. Nevertheless, this approach has proven to be useful in revealing general patterns of intersectional inequalities in the US (Kozlowski, Larivière, et al., 2022), and was also manually validated in our previous work (Kozlowski, Murray, et al., 2022). The relative overrepresentation seen for Latinx authors in HSIs and of Black authors in HBCU (see Fig. S5) also shows that the method holds face validity for the institutional level analysis. The main limitation of this model is, nevertheless, its inability to infer the demographics of Native American scholars. We are forced to omit this group from our analysis given this limitation, which reinforces the need for alternative methodologies that rely on self-identification of authors.

Gender was inferred using authors' given names and census data, based on the method presented in Larivière et al. (2013). In cases where a name was used for both genders, it was only attributed to a specific gender when it was used at least ten times more frequently for one gender than the other. Otherwise, it was categorized as a "unisex" name. Our previous validation showed a high precision for men and women. Nevertheless, given that the census information does not consider other genders, we could only consider gender in a binary way. This is a clear limitation of this algorithmic approach.

These limitations highlight the need for alternative work based on self-identification. It also shows the need for publishers and institutions to securely collect demographic information of authors for large-scale assessment of diversity in academia. Despite their limitations, the methods used in this paper shed a necessary light on institutional inequalities in academe and of their consequences.

Topics and indicators

The definition of fields used in this paper is based on a journal classification developed for the US NSF (Hamilton, 2003). The topic of articles is inferred using Latent Dirichlet Allocation models (LDA). Based on our previous work (see Kozlowski, Larivière, et al. (2022), we train a model for Social Science, Humanities and Professional Fields with 300 topics, and a model with 200 topics for each of the other fields (including Health). Optimizing hyperparameters, such as the number of topics, for an unsupervised method like LDA is not straightforward. We selected the number of topics based on a manual exploration aimed at providing detailed topics while avoiding repetition. To ensure robustness, we compared several runs of the LDA model with different seeds for the Social Science, Humanities and Professional Fields model, and evaluated the closeness of their predictions with respect to the health model and a random Dirichlet model (Kozlowski, Larivière, et al., 2022). For this dataset, we found the LDA model to be robust to different seeds. Despite using different numbers of topics, the manual exploration revealed repeated patterns, indicating the model's general consistency in explaining research topics. The topical alignment is based on the proportion of papers each group—race & gender identities or institutional groups—contributes to each topic. Race & gender are assigned probabilistically to each article and topic. To account for the proportion of papers that a researcher identity produces in a topic, we sum for all papers their probability associated to that topic multiplied by the probability of that paper being written by an author of that same identity. For institutional groups, which are categorical, we sum the probabilities associated with topics for each group separately, and then divide by the sum of probabilities for that topic across all groups. The result obtained is the proportion of papers each race, gender, and institutional group contributed to each topic. Then, the correlation between each group and participation in a given topic can be made for any institutional category. Fig. 2 shows the correlations between institutions and identities for the *perceived* prestige, and alternative prestige metrics are shown in the supplementary (see Fig S.10-S18).

As a complement to the results presented in this article, we deployed a website companion (https://sciencebias.ebsi.umontreal.ca/) with alternative metrics and dynamic visualizations. There, the labels of all topics can be hover over the scatterplots, which show the results for all fields. The correlations for all fields are also shown.

Topic-normalization and Linear models

To build the topic normalized citations and JIF, we divided the 2-year citations of articles by the average number of 2-year citations for their respective topic and normalized the JIF of articles by the average JIF of their respective topic. As each article has a probability distribution across topics, we used this distribution as weights for a weighted average to perform the normalization. Fig. S4 illustrates the difference between the topic-normalized approach and the field-normalized approach, which is commonly used in scientometric studies. By using topic-normalization directly on the dependent variable, we were able to build a model that includes all fields. Since each field has its own set of topics, making the inclusion of topics as covariates would be unfeasible.

With articles as unit of analysis, we built two types of models: First, the aggregated model that includes as covariables the race and gender identity, the institutional category, and the number of authors and career age (time between year of first publication and the year of the published article) as controls. The aggregated models show the following structure:

$$y = \beta_0 + \beta_1 \#authors + \beta_2 career\ age + \sum_i \beta_i\ race\ \&\ gender + \sum_j \beta_j\ institution$$

where $y$ is the year- and topic-normalized (or field-normalized) citations or JIF, $\#authors$ is the number of authors, $race\ \&\ gender$ is the first author probability of being from a specific race and gender group. Researchers' identities are computed as probabilities, and the sum of those probabilities adds to one. Therefore, we exclude the *White men*'s group to avoid multicollinearity. This means that all the other values should be understood as the effect of being from a specific group in comparison to the White man category. As we work with three proxies of prestige, we built a model for each of the categories described above: US News & World report, average citations, and Carnegie selectivity. In the analysis we splitted each of these categorizations into three groups. For the linear model, we omit the low prestige group, and therefore the medium and high prestige covariables are read as the effect with respect to the low prestige group. For instance, in Fig. 4, β = 0.57 for Top 10 institutions implies that an article from a Top 10 university is expected to receive 57% more citations than an article from a non-top institution from the same topic, while controlling for identity, career age, and number of authors. The average number of citations of the institution is a continuous variable by construction, so we also added a model to consider the continuous distribution.

However, these models do not permit us to observe how the impact gap varies across different institutional types. To address this, we constructed a second set of models by dividing the population into each of the groups for each of the prestige proxies. This resulted in 9 models: three models per institutional category. These models also control for both topical and institutional profile of authors, but in a different manner, with each model focusing on a particular subset of the author population. This enables us to compare, for instance, the impact gap for Black women in Top 10 institutions versus the impact gap for Black women in non-top institutions. The results of these models are presented in Fig. 5, and can be defined as follows:

$$y = \beta_{0j} + \beta_{1j} \#authors + \beta_{2j}\, career\ age + \sum_i \beta_{ij}\, race\ \&\ gender$$

where *j* is each of the three groups—low, middle and high prestige—of each of the institutional categorisations—US News & World report, average citations, and Carnegie selectivity index—.

In this model the $\beta$=-0.36 for Latinx women in the Top 10 institutions (see Fig. 5 and table S3) means that we expect that a Black women author from a Top 10 institution will receive on average 36% less citations than an White men from a Top 10 institution for an article in the same topic, after controlling for number of authors and career age, while the $\beta$=-0.06 for the non top institutions shows that the expected difference is "only" 6% among authors from those institutions.

For all models, the level of analysis are articles. As we consider only the first authors, each article can be associated with a distribution of probability over race and gender, another distribution of probability over topics (used for normalization), and the institution of belonging of that first author. Using articles as a unit of analysis avoids the problem of authors disambiguation.

Alternatively, we also build an interaction model between race and gender, which shows consistent results with the ones presented in this article. Results for this model are showed in https://sciencebias.ebsi.umontreal.ca/.

**References for the Supplementary material**

Carnegie. (2022). *Carnegie Classifications | Undergraduate Profile Classification*. https://carnegieclassifications.acenet.edu/classification_descriptions/undergraduate_profile.php

Furstenberg, F. (2007). *In the Name of the Father: Washington's Legacy, Slavery, and the Making of a Nation* (Illustrated edition). Penguin Books.

Hamilton, K. S. (2003). *Subfield and level classification of journals* (CHI Research No. 2012-R).

Kozlowski, D., Larivière, V., Sugimoto, C. R., & Monroe-White, T. (2022). Intersectional inequalities in science. *Proceedings of the National Academy of Sciences*, *119*(2), e2113067119. https://doi.org/10.1073/pnas.2113067119

Kozlowski, D., Murray, D. S., Bell, A., Hulsey, W., Larivière, V., Monroe-White, T., & Sugimoto, C. R. (2022). Avoiding bias when inferring race using name-based approaches. *PLOS ONE*, *17*(3), e0264270. https://doi.org/10.1371/journal.pone.0264270

Laberge, N., Wapman, K. H., Morgan, A. C., Zhang, S., Larremore, D. B., & Clauset, A. (2022). Subfield prestige and gender inequality among U.S. computing faculty. *Communications of the ACM*, *65*(12), 46–55. https://doi.org/10.1145/3535510

Larivière, V., Ni, C., Gingras, Y., Cronin, B., & Sugimoto, C. R. (2013). Bibliometrics: Global gender disparities in science. *Nature*, *504*(7479), Article 7479. https://doi.org/10.1038/504211a

Tzioumis, K. (2018). Demographic aspects of first names. *Scientific Data*, *5*(1), Article 1. https://doi.org/10.1038/sdata.2018.25

USBC, U. C. B. (2016). *Frequently Occurring Surnames from the 2010 Census*. Census.Gov. https://www.census.gov/topics/population/genealogy/data/2010_surnames.html

Waltman, L., & van Eck, N. J. (2019). Field Normalization of Scientometric Indicators. In W. Glänzel, H. F. Moed, U. Schmoch, & M. Thelwall (Eds.), *Springer Handbook of Science and Technology Indicators* (pp. 281–300). Springer International Publishing. https://doi.org/10.1007/978-3-030-02511-3_11

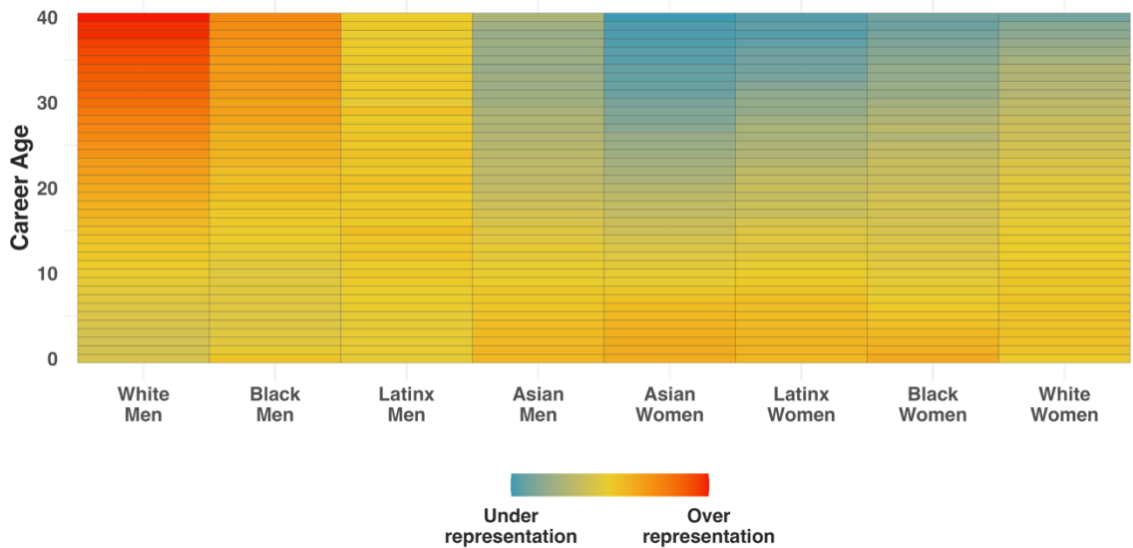

**Fig. S1. Relative under and overrepresentation by identity and career age.** For all authors with more than one publication, given the latest publication of each author in the period 2008-2020, including authorship in non-first position.

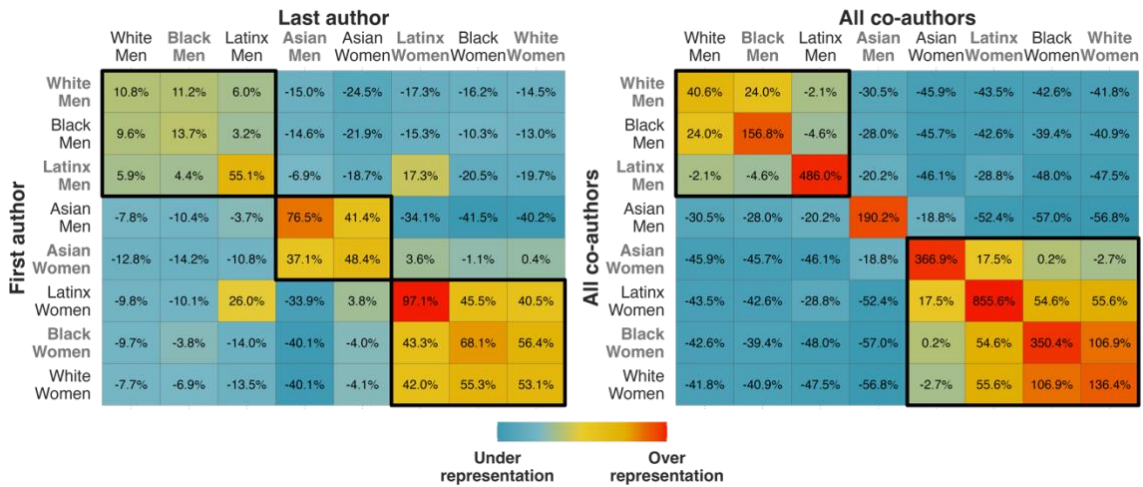

**Fig. S2. Over and underrepresentation of authors collaborations by race and gender.** The figure on the left shows the relation between first and last authors, while the figure on the right shows the relation between all co-authors. Both cases are computed on the US articles between 2008 and 2019 (5,431,451 articles). The values represent the proportion of cases of more (or less) with respect to the randomly expected value, controlled by the disciplinary distribution by race and gender over fields.

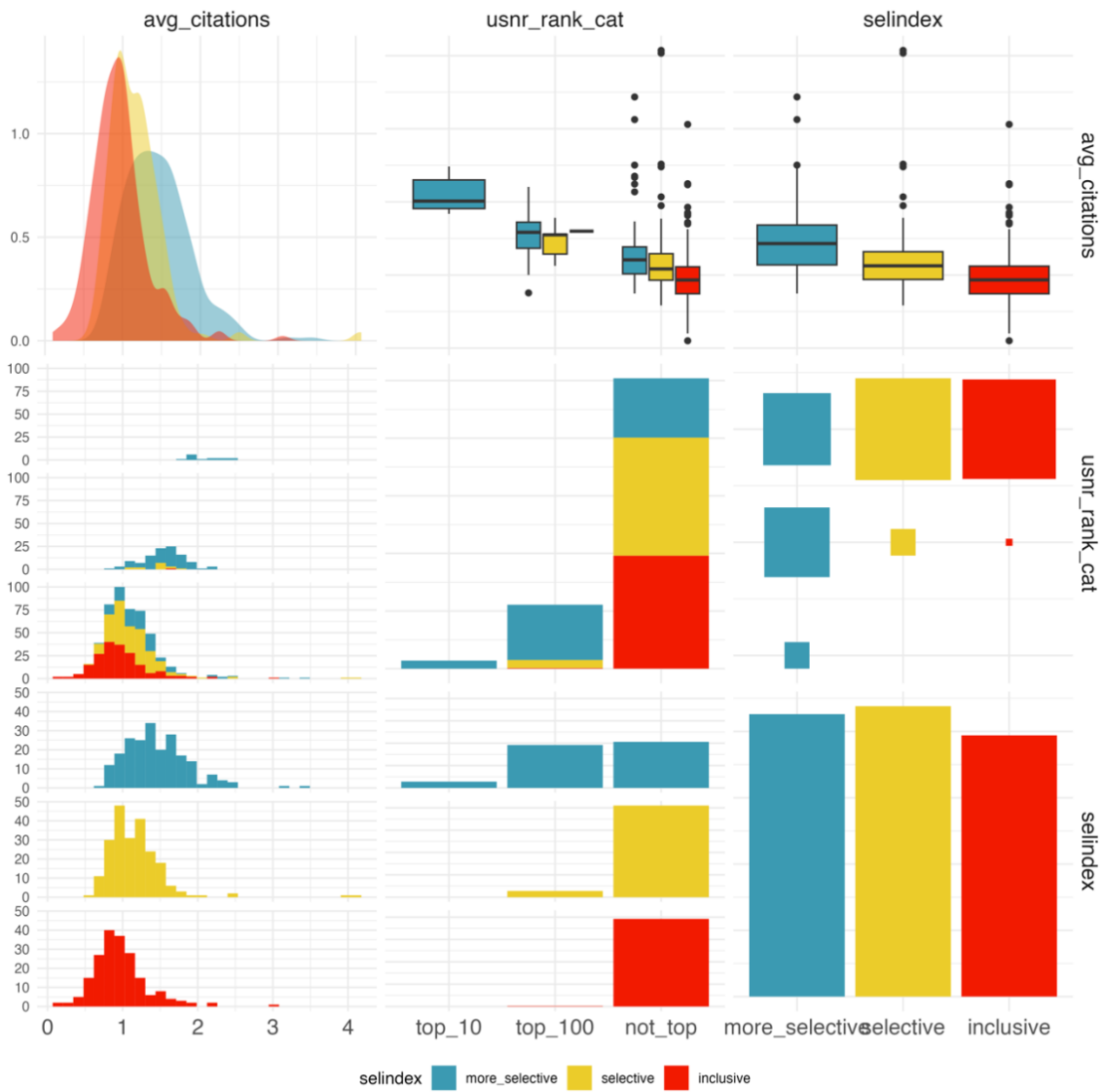

**Fig. S3. Prestige metrics.** Relation between the three proxies of prestige. We can see that more selective institutions (selindex) have also more citations on average (A., C. and G.), that top 10 institutions (US News & World report) are only on the more selective group (selindex) (B. and F.), and have more citations than the top 100 and those not in the top (B.), most inclusive and selective institutions are outside the top 100 of US News and World report (F. and H.), and within this group, we can still see that more selective institutions also have more citations (B.). The number of institutions is well balanced in the Selectivity index (I.) but not in the US News & World report groups, although this also means, given the productivity differences, that the number of authors and articles is more balanced on the US News & World report than on the selectivity index, while the groups build from the historical number of citations are built to balance the number of articles (see Table S1)

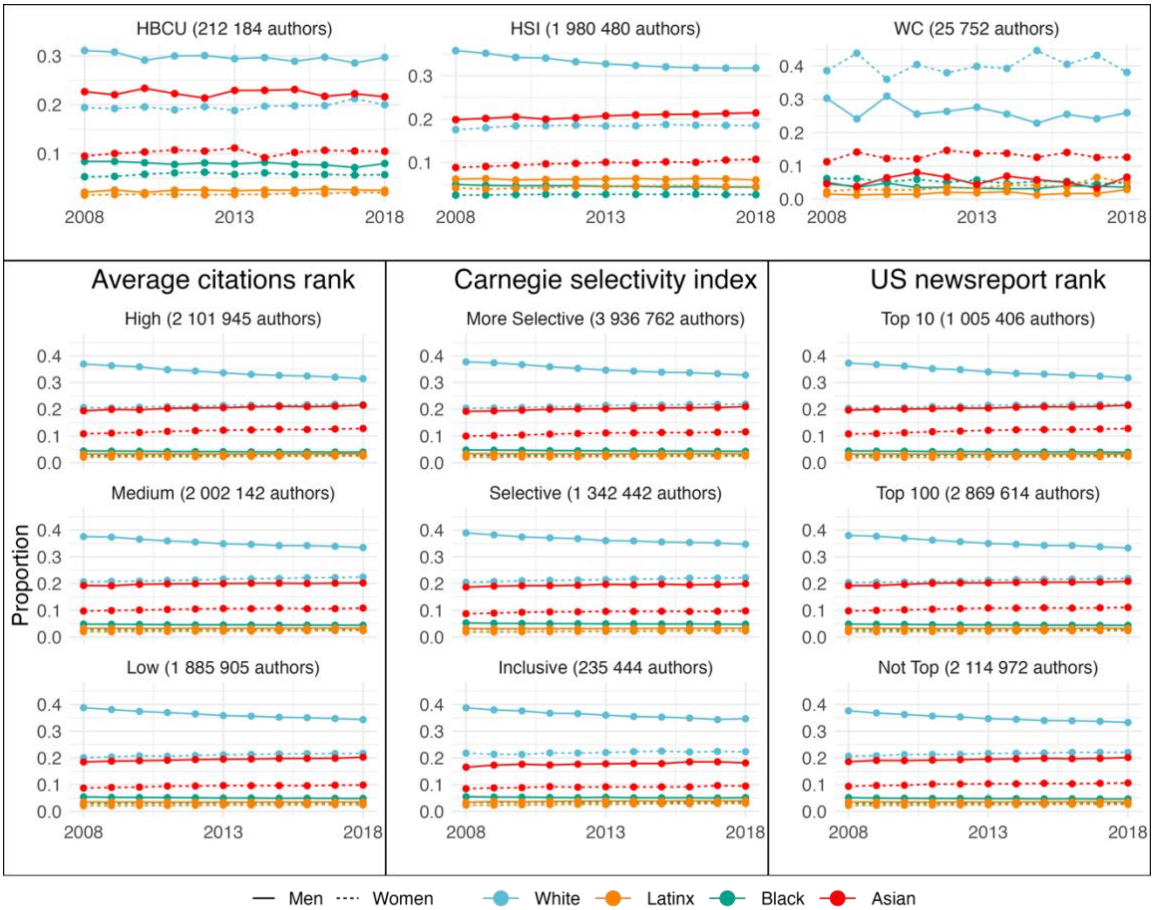

**Fig. S4. White men are still largely overrepresented with respect to their proportion in the US Census across all institution types.** Proportion of groups by race and gender, for the number of authors. HBCU: Historically Black Colleges and Universities, HSI: Hispanic Serving Institutions, and WC: Women's colleges. Institutions sorted by their average number of citations: Low (0.1, 1.47), Medium (1.48, 1.74), and High (1.77, 4.07). Carnegie Selectivity Index based on admissions rates. US News & World ranking: Top 10 institutions, Top 100 institutions (without the Top 10) and institutions not in the Top 100. Total number of authors between parentheses.

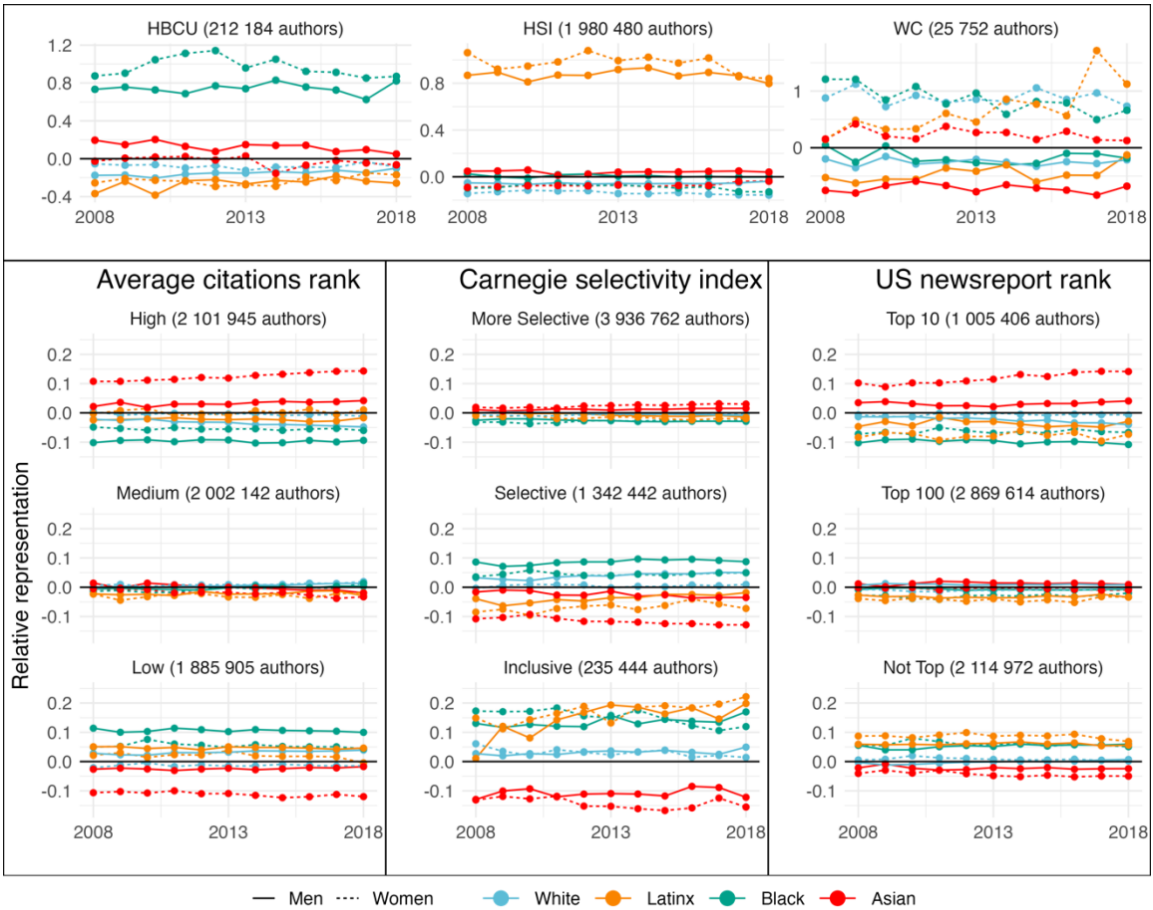

**Fig. S5 Institutions serving specific groups show a larger authorship from those groups, while low prestige institutions show a larger proportion of Black and Latinx authors.** Relative over/under representation of groups by race and gender, relative to their participation in the overall dataset. HBCU: Historically Black Colleges and Universities, HSI: Hispanic Serving Institutions, and WC: Women's colleges. Institutions sorted by their average number of citations: Low (0.1, 1.47), Medium (1.48, 1.74), and High (1.77, 4.07). Carnegie Selectivity Index based on admissions rates. US News & World Report ranking: Top 10 institutions, Top 100 institutions (without the Top 10) and institutions not in the Top 100. Total number of authors between parentheses. Women and minority serving institutions:

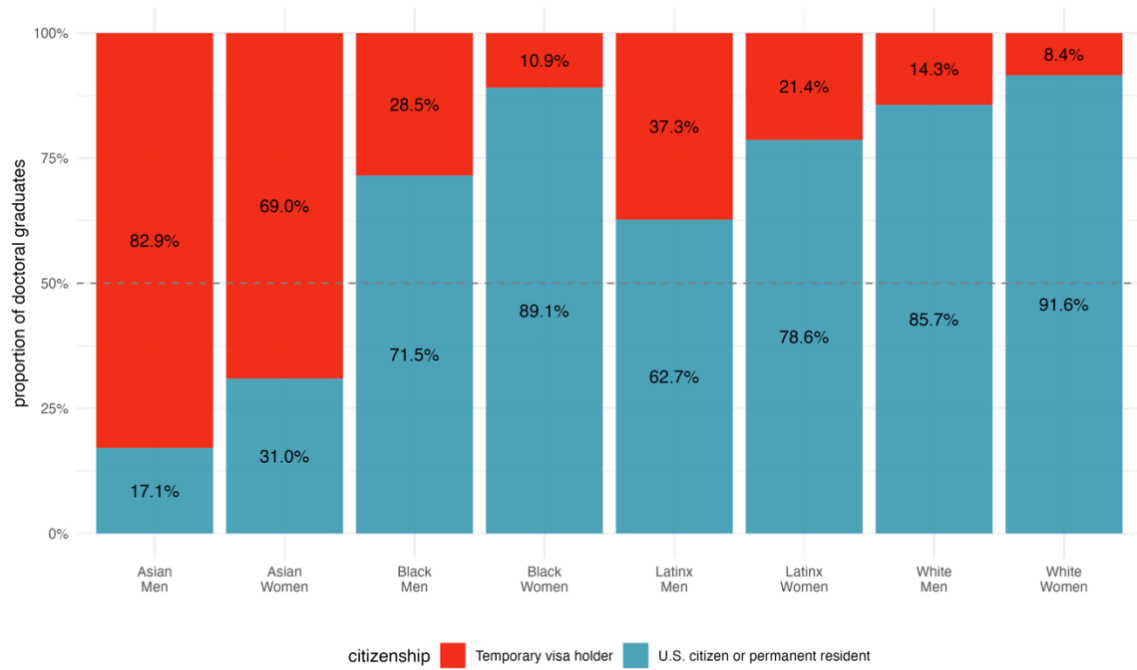

**Fig. S6. Migration patterns.** Proportion of doctoral graduates that are temporary visa holders and U.S. citizens or permanent residents, according to Survey of Earned Doctorates.

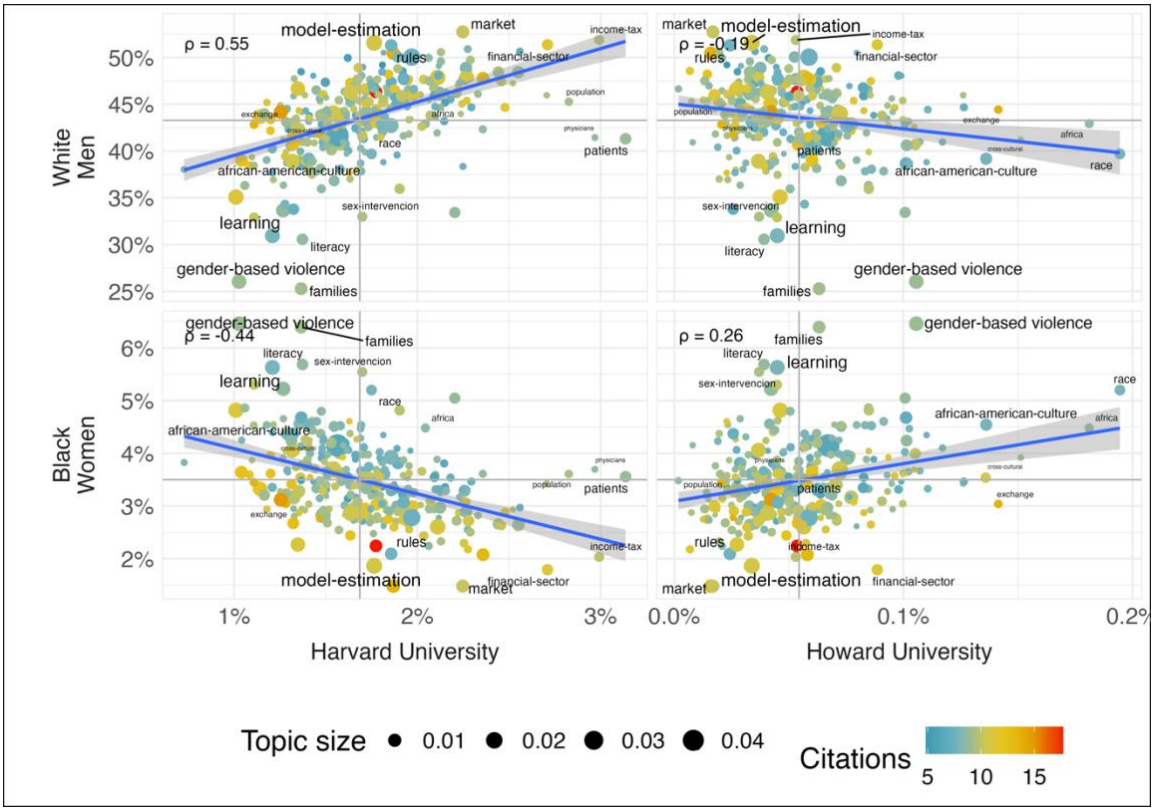

**Fig. S7. Topical alignment of institutions and identities.** Proportion of papers in different topics authored by Black Women and White Men (vertical axis) and the percentage of those papers authored by two exemplar institutions: Howard University and Harvard University. Dot size represents the size of the topic in the corpus associated with the topic, while the dot color represents the average number of citations for that topic. For each subplot, ρ indicates the Spearman correlation, and the blue line is the simple linear regression between the two variables, with the 95% confidence interval. The top 5 topics of each identity and institution group were hand labelled.

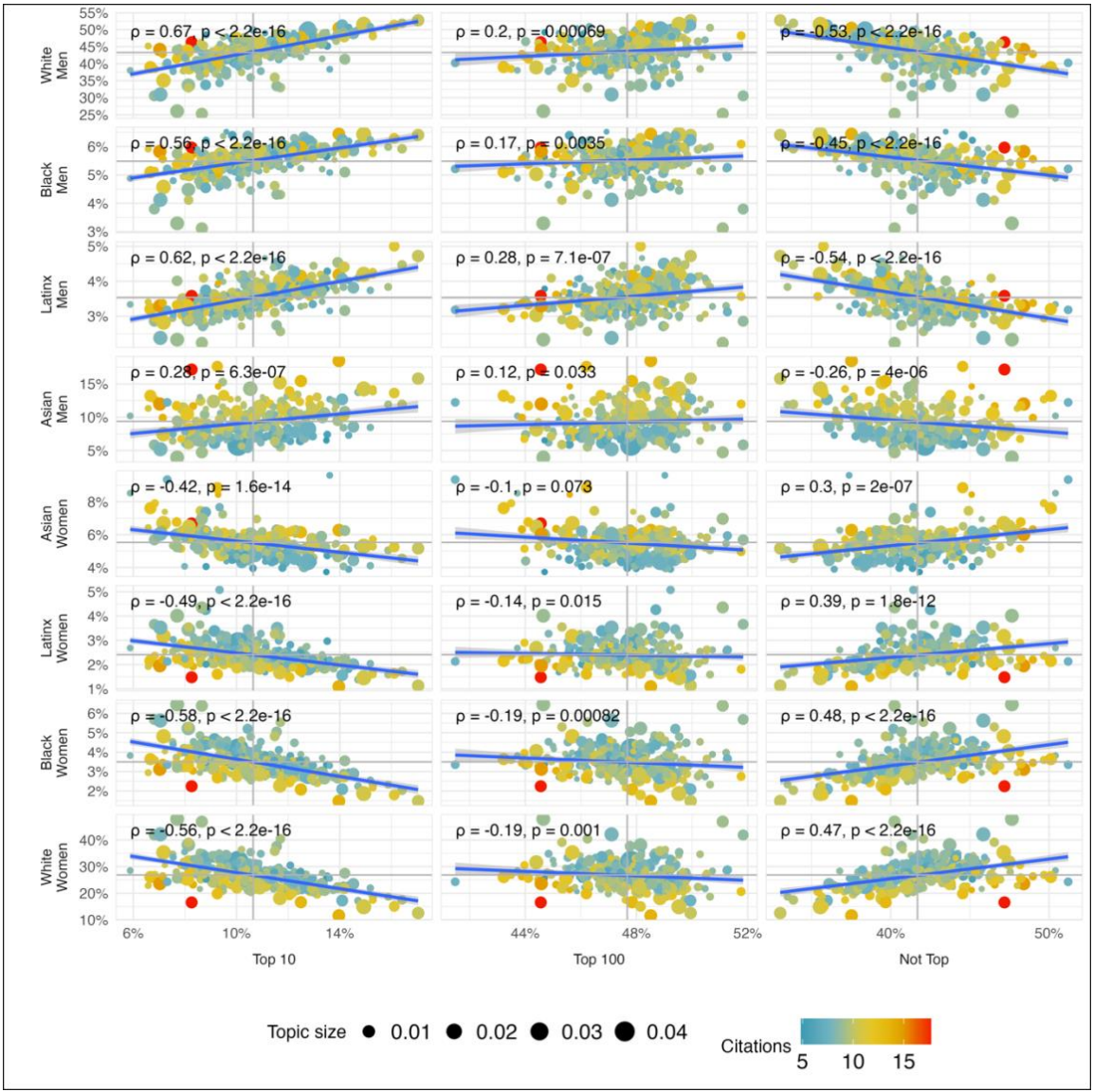

**Fig. S8. Relationship between topic representation of authors by race and gender, and topic representation of institutional groups, for papers in the Social Sciences, Humanities and Professional Fields.** The figures provide Spearman correlations between the proportion of papers in different topics authored by race and gender (vertical axis) and three categories of perceived prestige (horizontal axis) based on the US News and World report ranking: Top 10 institutions, Top 100 institutions (without the Top 10) and institutions not in the Top 100. Dot size represents the size of the topic in the corpus associated with the topic, while the dot color represents the average number of citations for that topic. For each subplot, ρ indicates the Spearman correlation with its p-value, and the blue line is the simple linear regression between the two variables. The gray area around the blue line represents the 95% confidence interval.

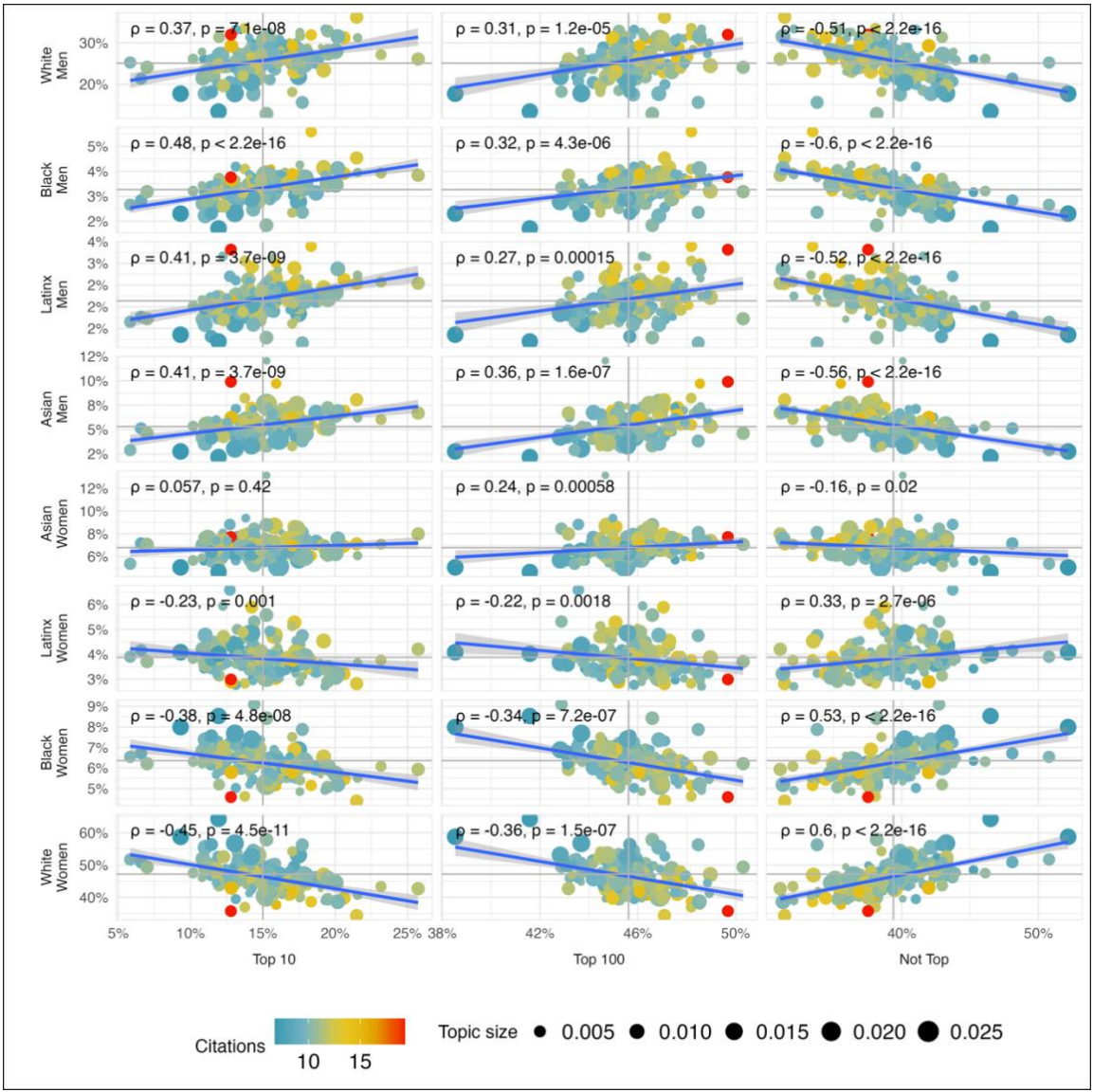

**Fig. S9. Relationship between topic representation of authors by race and gender, and topic representation of institutional groups, for papers in Health.** The figures provide Spearman correlations between the proportion of papers in different topics authored by race and gender (vertical axis) and three categories of perceived prestige (horizontal axis) based on the US News and World report ranking: Top 10 institutions, Top 100 institutions (without the Top 10) and institutions not in the Top 100. Dot size represents the size of the topic in the corpus associated with the topic, while the dot color represents the average number of citations for that topic. For each subplot, ρ indicates the Spearman correlation with its p-value, and the blue line is the simple linear regression between the two variables. The gray area around the blue line represents the 95% confidence interval.

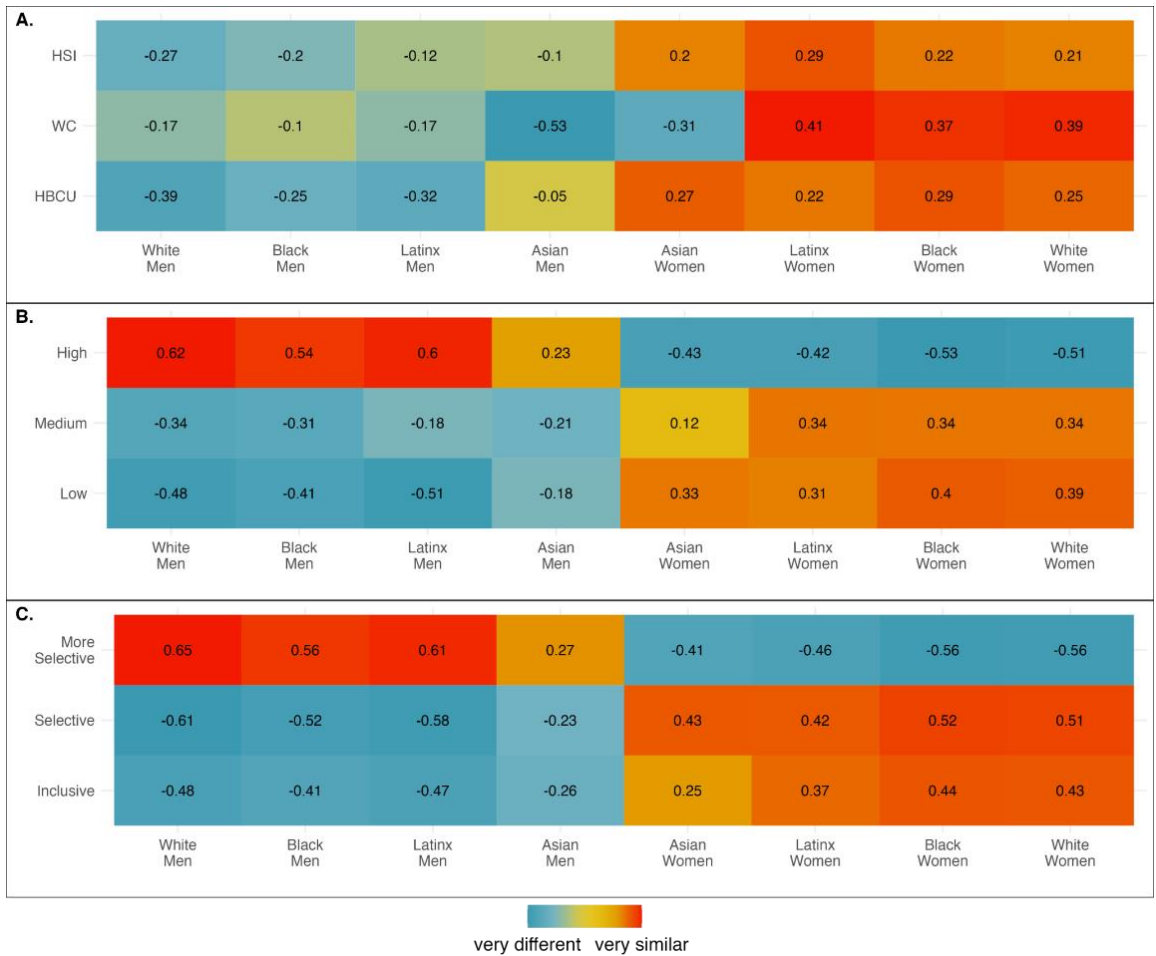

Figure S10. Spearman correlations between the topic profiles of each author identity and the topical profile of institutional categories for Social Sciences, Humanities and Professional Fields. Panel (A) provides correlations for institutions that serve specific groups: Historically Black Colleges and Universities (HBCU), Hispanic Serving Institutions (HSI), and Women's colleges (WC). Panel (B) provides correlations according to institutions ranked by their average number of citations (*Research* prestige): Low (0.1, 1.47), Medium (1.48, 1.74), and High (1.77, 4.07); and Panel (C) provides correlations according to institutions according to *Selectivity* prestige: Carnegie Selectivity Index based on admissions rates.

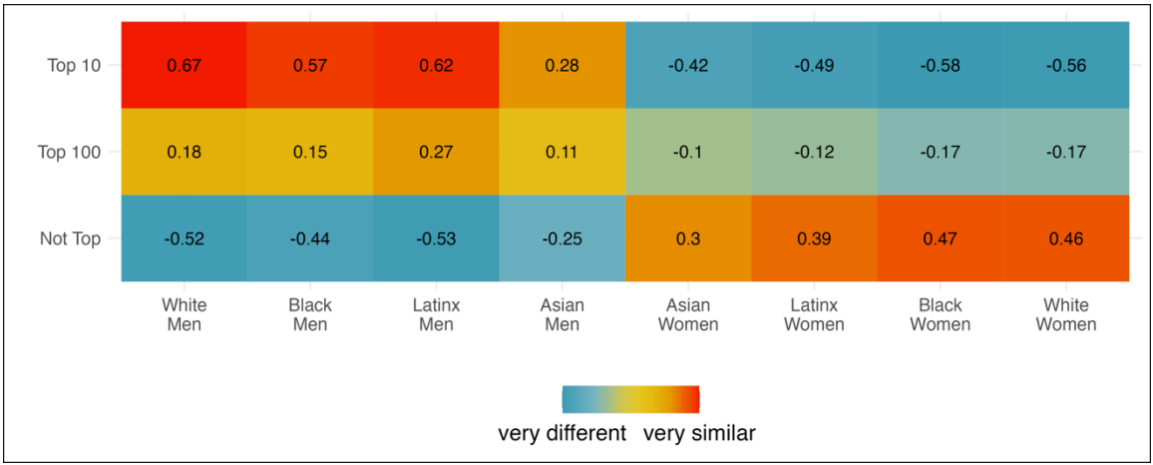

**Figure S11. Correlations corrected by compositional effect.** Spearman correlations between the topical profiles of each author identity and the ratio between the actual and expected topical profiles of the institutional group given the race and gender distribution of authors. For Social Sciences, Humanities and Professional Fields. Institutions divided according to *Perceived* prestige from US News & World Report: Top 10 institutions, Top 100 institutions (without the Top 10), and institutions not in the Top 100.

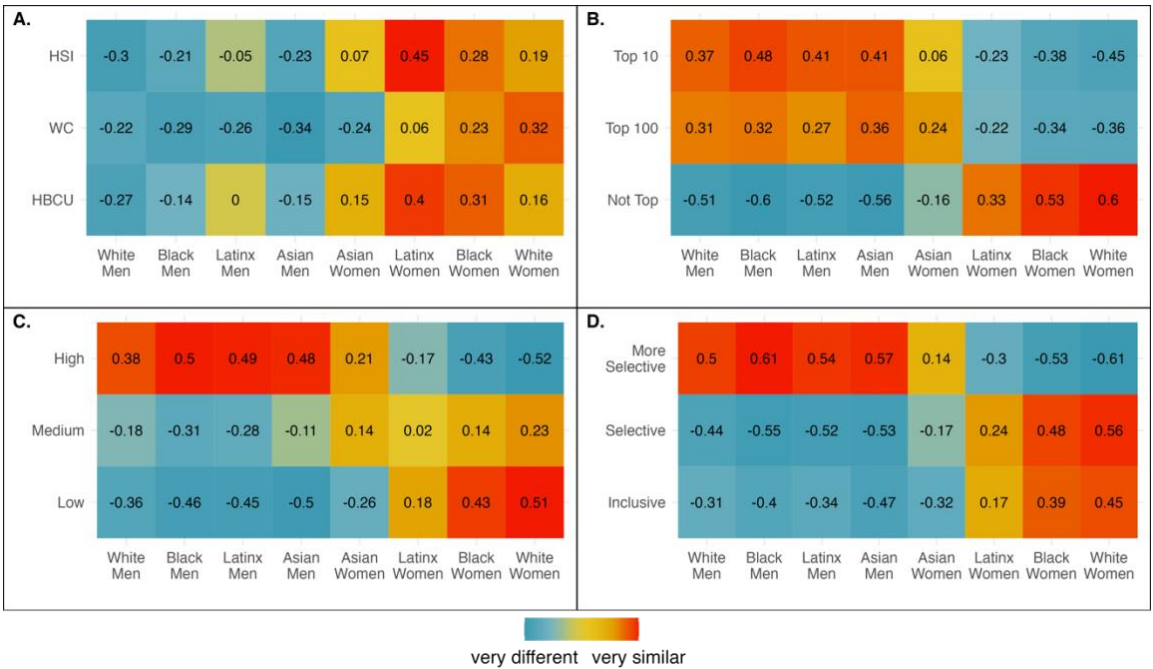

**Fig. S12. Spearman correlations between the topic profiles of each author identity and the topical profile of institutional categories for Health.** Panel **(A)** provides correlations for institutions that serve specific groups: Historically Black Colleges and Universities (HBCU), Hispanic Serving Institutions (HSI), and Women's colleges (WC). Panel **(B)** provides correlations for institutions divided according to *Perceived* prestige from US News & World Report: Top 10 institutions, Top 100 institutions (without the Top 10), and institutions not in the Top 100. Panel **(C)** provides correlations according to institutions ranked by their average number of citations (*Research* prestige): Low (0.1, 1.47), Medium (1.48, 1.74), and High (1.77, 4.07); and Panel **(D)** provides correlations according to institutions according to *Selectivity* prestige: Carnegie Selectivity Index based on admissions rates.

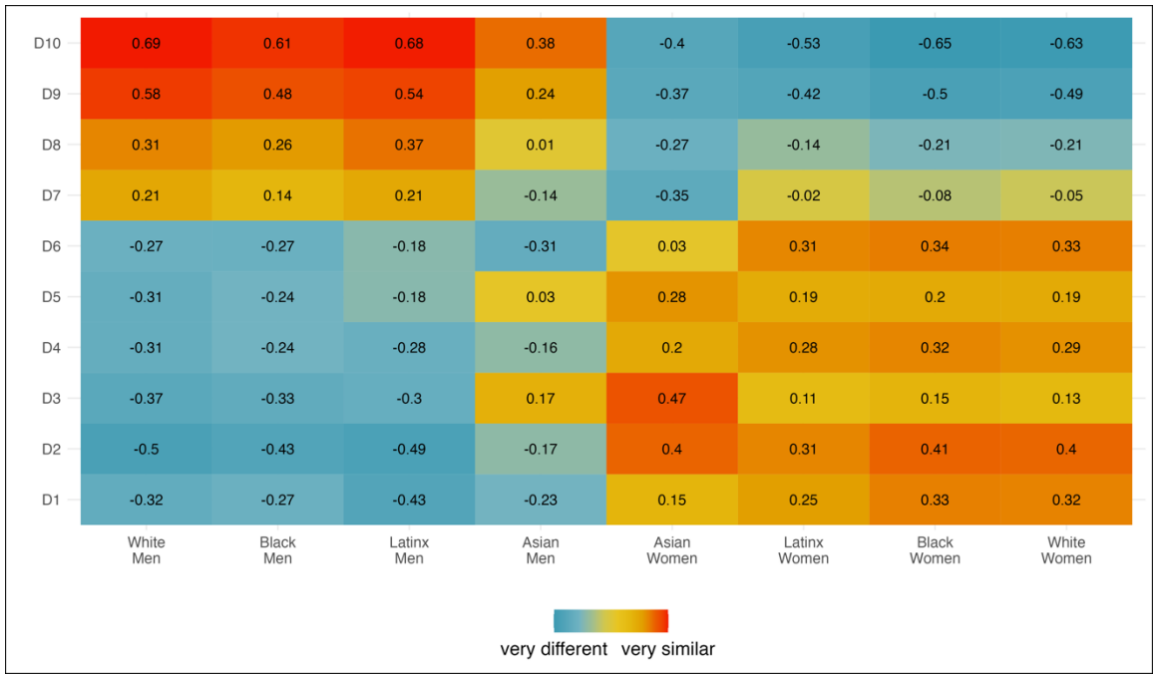

**Fig. S13. Spearman correlations between the topic profiles of each author identity and the topical profile of institutional categories for Social Sciences, Humanities and Professional Fields.** Institutions are sorted by their average number of citations (*Research* prestige) into deciles, from the most cited (D10) to the least cited (D1). This figure presents a more granular representation of the Low, Medium and Highly cited groups for research prestige.

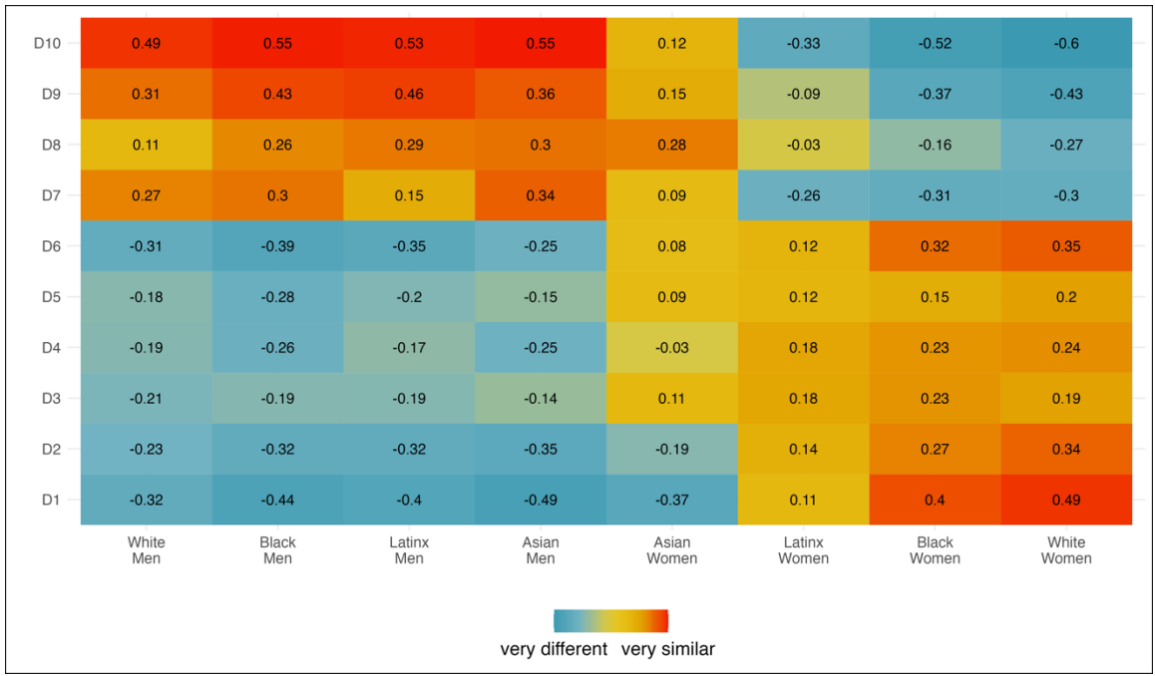

**Fig. S14. Spearman correlations between the topic profiles of each author identity and the topical profile of institutional categories for Health.** Institutions are sorted by their average number of citations (*Research* prestige) into deciles, from the most cited (D10) to the least cited (D1). This figure presents a more granular representation of the Low, Medium and Highly cited groups for research prestige.

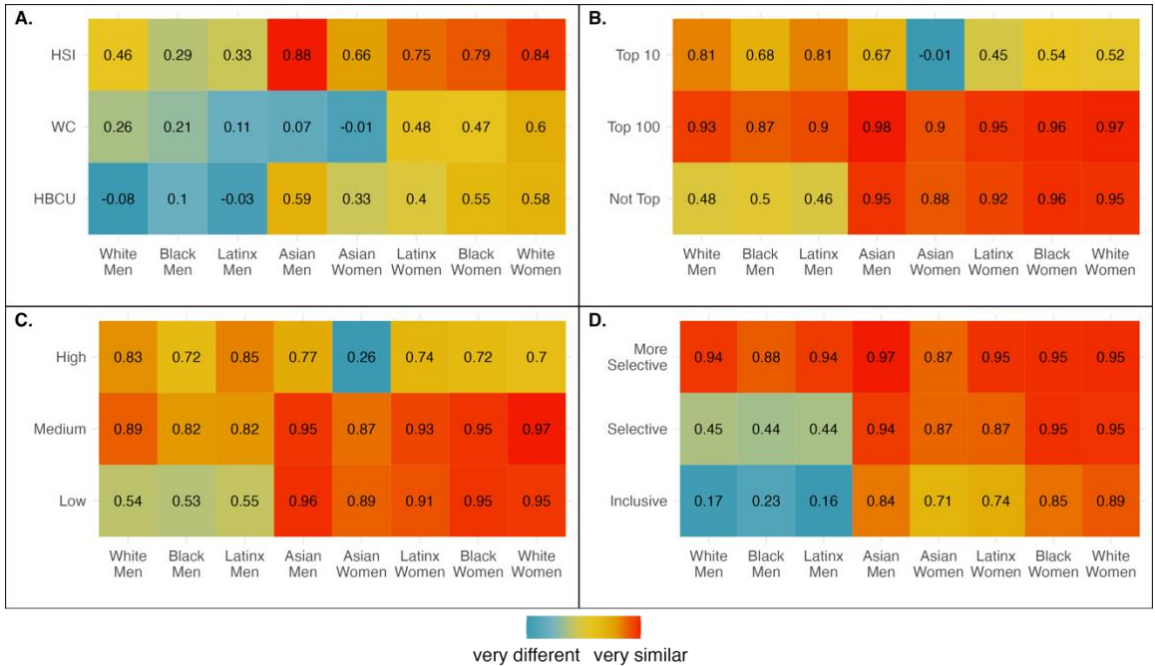

**Fig. S15. Spearman correlations between the topic profile of each author identity within an institutional category and the topic profiles of each author identity across all institutional categories for Social Sciences, Humanities and Professional Fields.** Panel **(A)** provides correlations for institutions that serve specific groups: Historically Black Colleges and Universities (HBCU), Hispanic Serving Institutions (HSI), and Women's colleges (WC). Panel **(B)** provides correlations for institutions divided according to *Perceived* prestige from US News & World Report: Top 10 institutions, Top 100 institutions (without the Top 10), and institutions not in the Top 100. Panel **(C)** provides correlations according to institutions ranked by their average number of citations (*Research* prestige): Low (0.1, 1.47), Medium (1.48, 1.74), and High (1.77, 4.07); and Panel **(D)** provides correlations according to institutions according to *Selectivity* prestige: Carnegie Selectivity Index based on admissions rates.

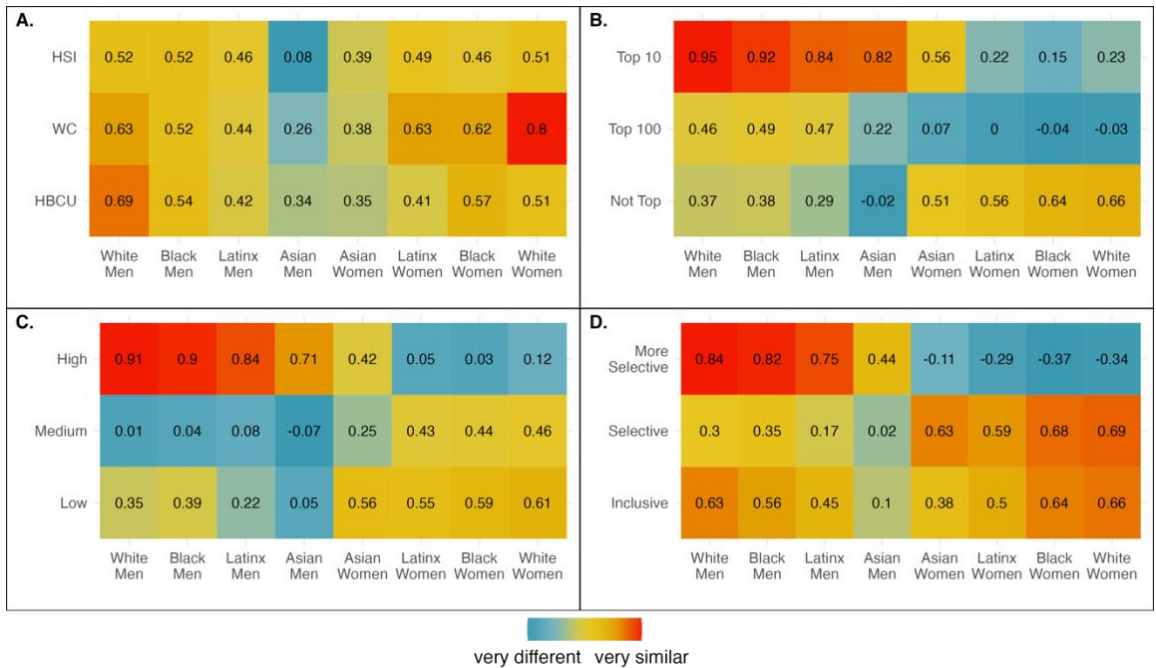

**Fig. S16. Spearman correlations between the topic profile of each author identity within an institutional category and the topic profiles of all authors from that institutional category for Social Sciences, Humanities and Professional Fields.** Panel **(A)** provides correlations for institutions that serve specific groups: Historically Black Colleges and Universities (HBCU), Hispanic Serving Institutions (HSI), and Women's colleges (WC). Panel **(B)** provides correlations for institutions divided according to *Perceived* prestige from US News & World Report: Top 10 institutions, Top 100 institutions (without the Top 10), and institutions not in the Top 100. Panel **(C)** provides correlations according to institutions ranked by their average number of citations (*Research* prestige): Low (0.1, 1.47), Medium (1.48, 1.74), and High (1.77, 4.07); and Panel **(D)** provides correlations according to institutions according to *Selectivity* prestige: Carnegie Selectivity Index based on admissions rates.

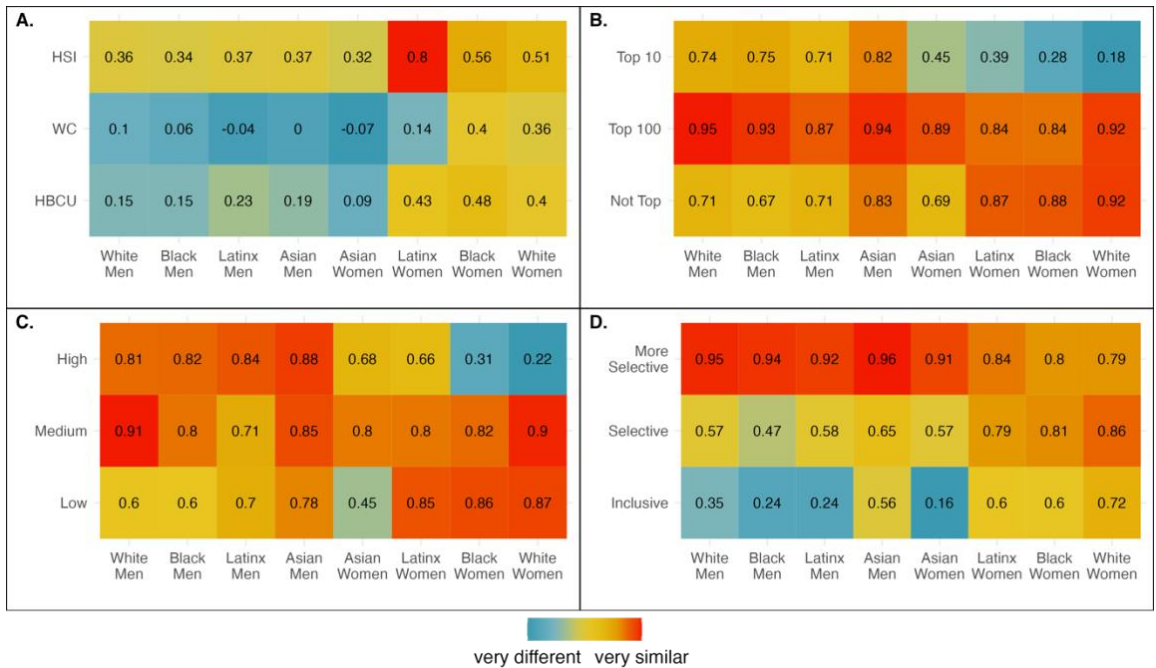

**Fig. S17. Spearman correlations between the topic profile of each author identity within an institutional category and the topic profiles of each author identity across all institutional categories for Health.** Panel **(A)** provides correlations for institutions that serve specific groups: Historically Black Colleges and Universities (HBCU), Hispanic Serving Institutions (HSI), and Women's colleges (WC). Panel **(B)** provides correlations for institutions divided according to *Perceived* prestige from US News & World Report: Top 10 institutions, Top 100 institutions (without the Top 10), and institutions not in the Top 100. Panel **(C)** provides correlations according to institutions ranked by their average number of citations (*Research* prestige): Low (0.1, 1.47), Medium (1.48, 1.74), and High (1.77, 4.07); and Panel **(D)** provides correlations according to institutions according to *Selectivity* prestige: Carnegie Selectivity Index based on admissions rates.

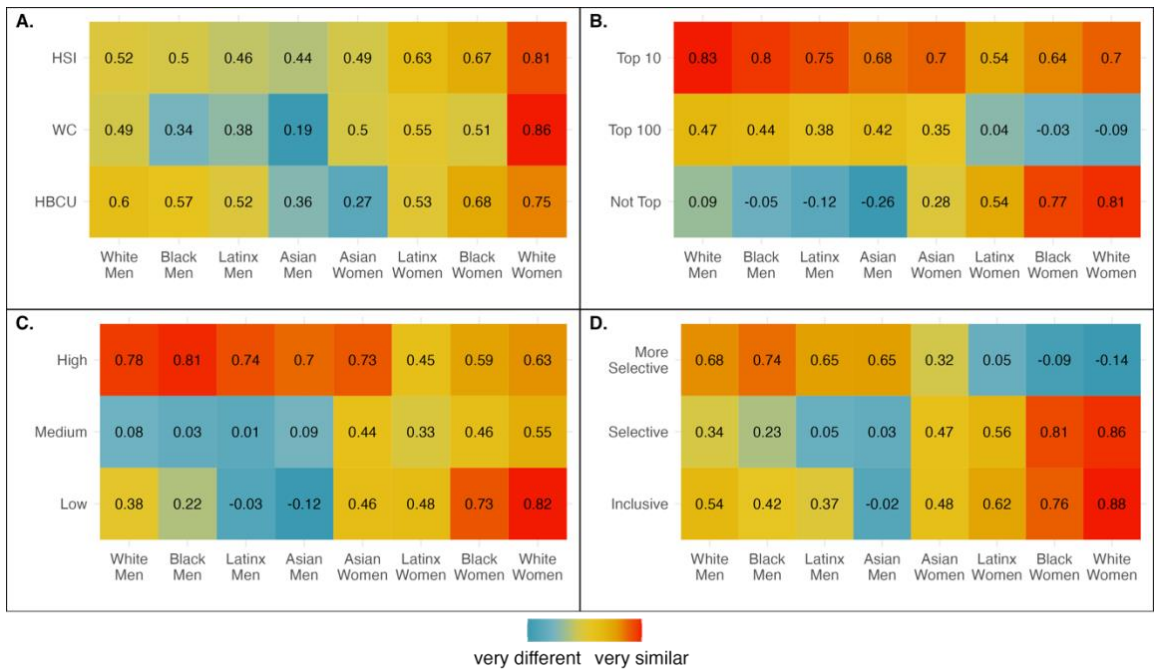

**Fig. S18. Spearman correlations between the topic profile of each author identity within an institutional category and the topic profiles of all authors from that institutional category for Health.** Panel **(A)** provides correlations for institutions that serve specific groups: Historically Black Colleges and Universities (HBCU), Hispanic Serving Institutions (HSI), and Women's colleges (WC). Panel **(B)** provides correlations for institutions divided according to *Perceived* prestige from US News & World Report: Top 10 institutions, Top 100 institutions (without the Top 10), and institutions not in the Top 100. Panel **(C)** provides correlations according to institutions ranked by their average number of citations (*Research* prestige): Low (0.1, 1.47), Medium (1.48, 1.74), and High (1.77, 4.07); and Panel **(D)** provides correlations according to institutions according to *Selectivity* prestige: Carnegie Selectivity Index based on admissions rates.

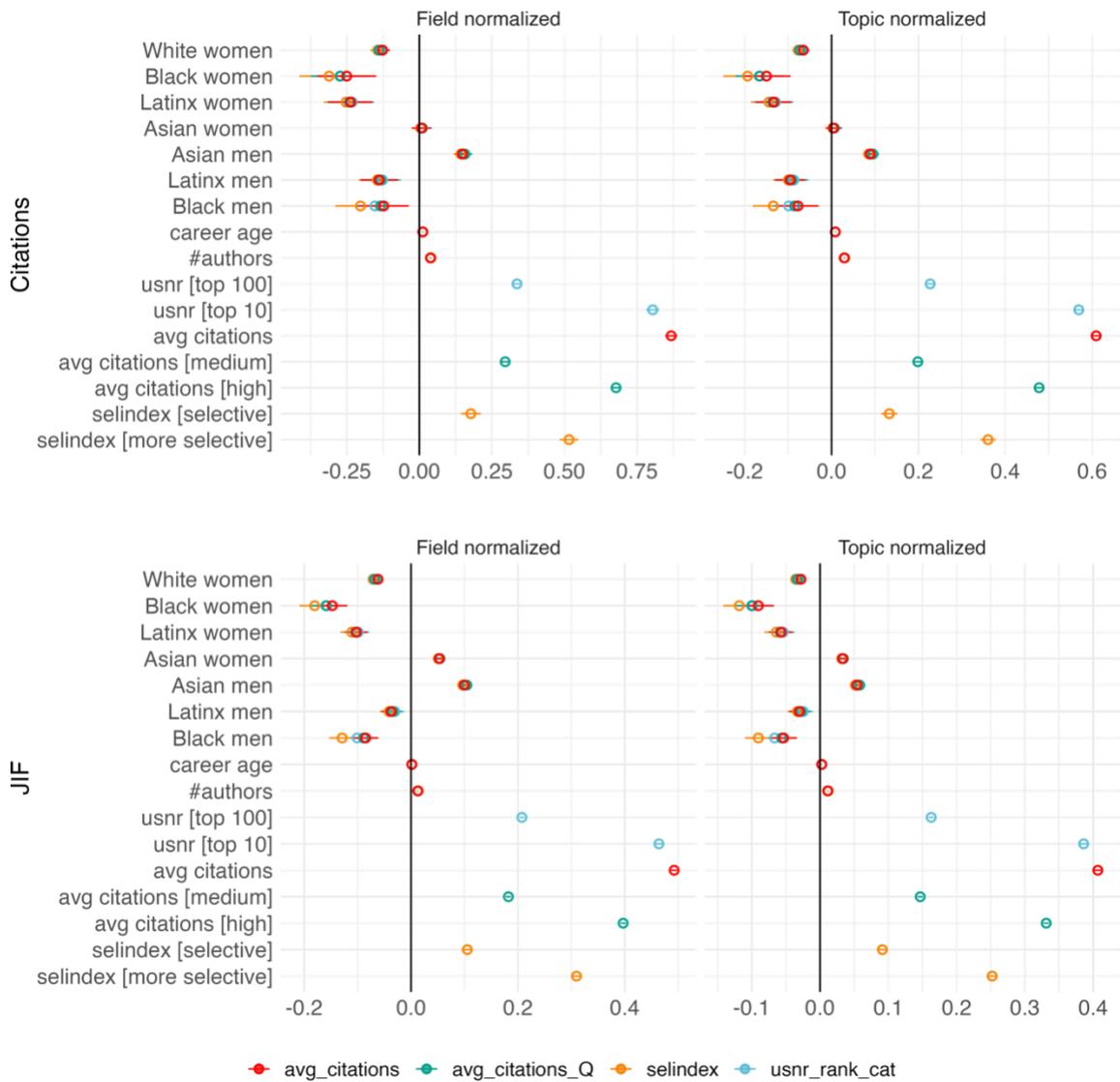

**Fig. S19. Parameters of linear regression models predicting citations and JIF, with and without topic normalization.** The reference group for our intersectional race by gender identity variables is White men, with the number of co-authors and career age serving as controls. Parameters of linear regression models predicting the two-year citations and JIF both with and without topic normalization. The unnormalized models scale the dependent variables (citations and JIF) by the average over the full dataset, while the normalized models scale the dependent variables by the average of the topic. The normalized version controls the effect of topics on impact. Each model was run with a different prestige indicator: *perceived* (US News & World Report): Top 10 institutions, Top 100 institutions (without the Top 10) and institutions not in the Top 100; *research* (institutions' historical average number of citations, both as a continuous, and categorical variable: Low (0.1, 1.47), Medium (1.48, 1.74), and High (1.77, 4.07); and *selective* (Carnegie Selectivity Index which is based on undergraduate admissions rates).

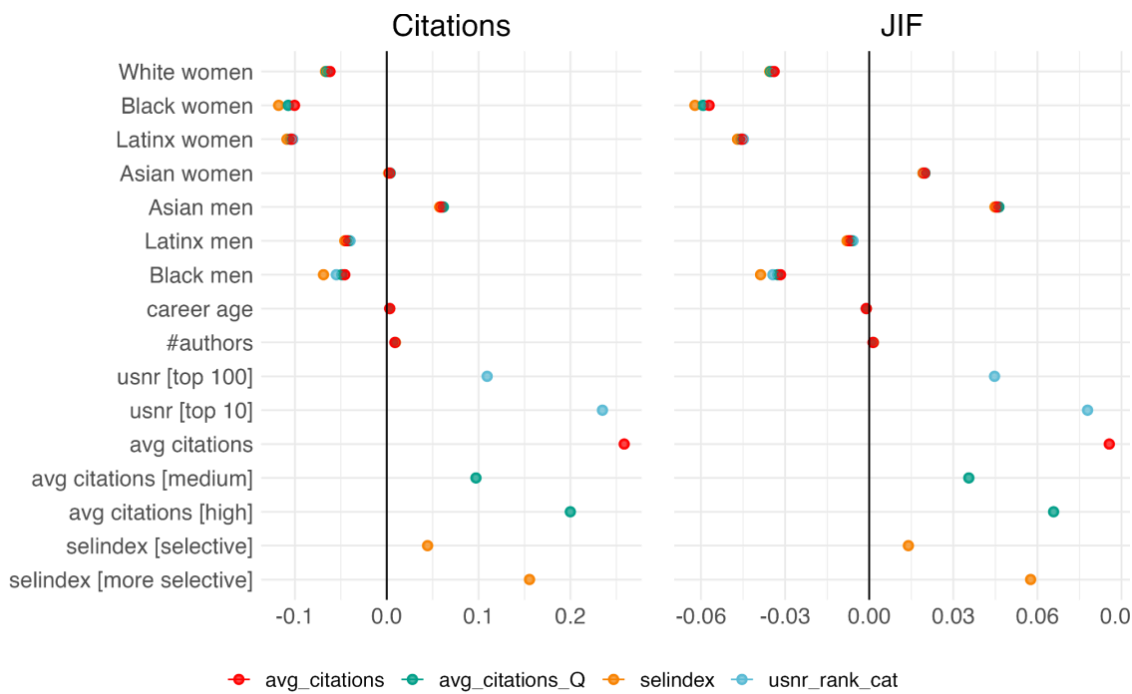

**Fig. S20. Difference between the parameters of linear regression models predicting citations and JIF, with and without topic normalization.** The reference group for our intersectional race by gender identity variables is White men, with the number of co-authors and career age serving as controls. Parameters of linear regression models predicting the two-year citations and JIF both with and without topic normalization. The unnormalized models scale the dependent variables (citations and JIF) by the average over the full dataset, while the normalized models scale the dependent variables by the average of the topic. The normalized version controls the effect of topics on impact. Each model was run with a different prestige indicator: *perceived* (US News & World Report): Top 10 institutions, Top 100 institutions (without the Top 10) and institutions not in the Top 100; *research* (institutions' historical average number of citations, both as a continuous, and categorical variable: Low (0.1, 1.47), Medium (1.48, 1.74), and High (1.77, 4.07); and *selective* (Carnegie Selectivity Index which is based on undergraduate admissions rates).

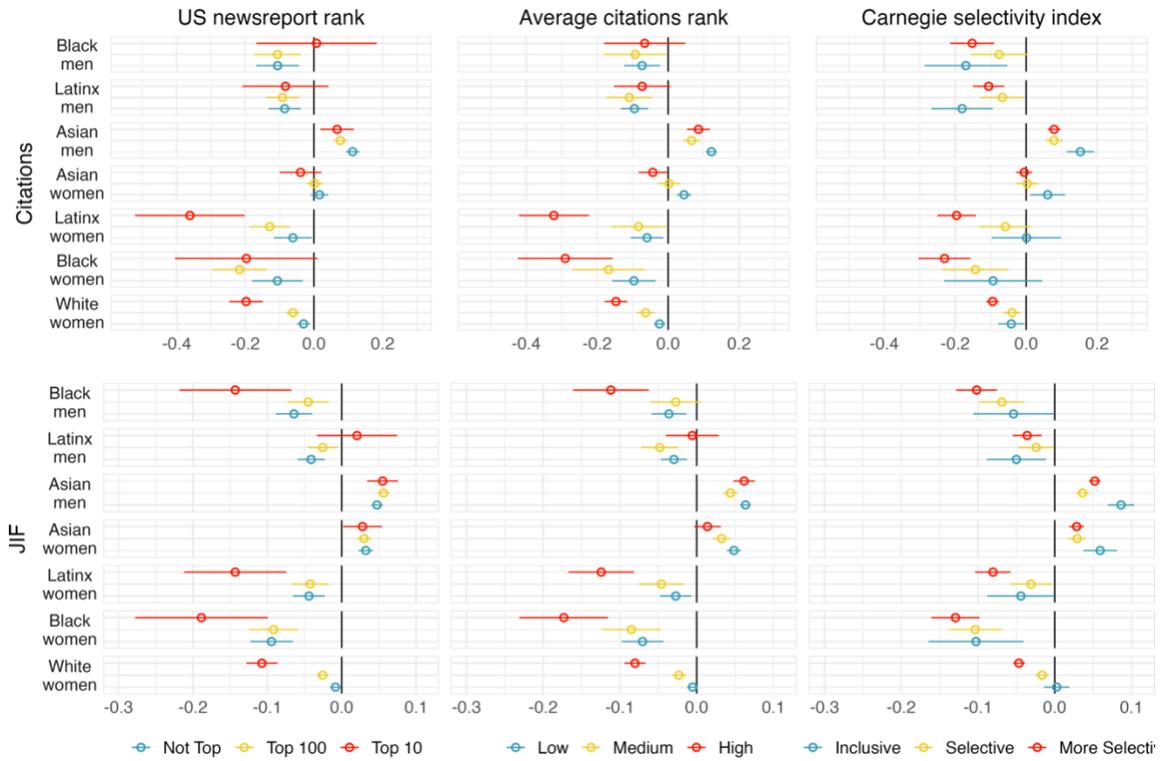

**Figure S21. Parameters of linear regression models predicting the topic and year normalized citations and JIF, for subsets of institutions**. The reference group for our intersectional race and gender identity variables is White men, with the number of co-authors and career age serving as controls. Each model was run with a different prestige indicator: *perceived* (US News & World Report): Top 10 institutions, Top 100 institutions (without the Top 10) and institutions not in the Top 100; *research* (institutions' historical average number of citations, both as a continuous, and categorical variable: Low (0.1, 1.47), Medium (1.48, 1.74), and High (1.77, 4.07); and *selective* (Carnegie Selectivity Index which is based on undergraduate admissions rates).

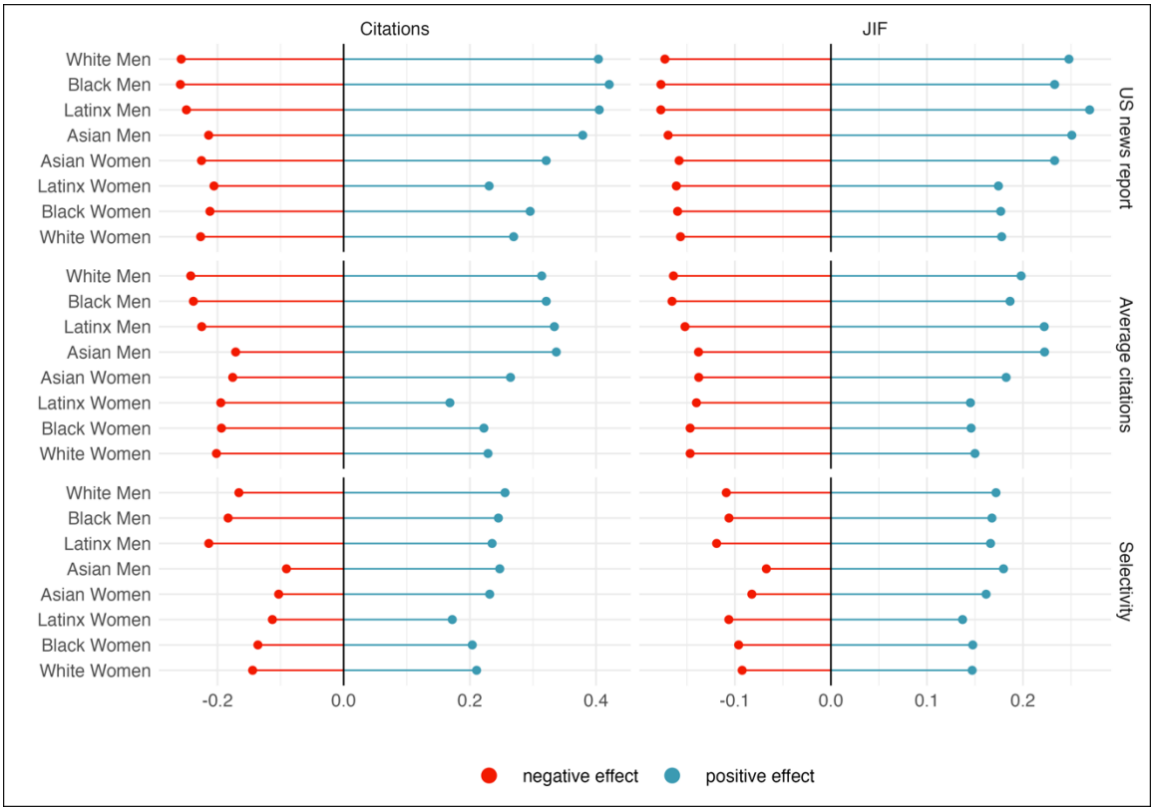

**Fig. S22. Negative and positive effects on citations and JIF, by race and gender between, across institutional groups.** For each institutional classification, we compute the increase in citations and JIF between the top and middle group, and the decrease in citations and JIF between the low prestige and middle group. Citations and JIF are topic and year normalized**.** US News & World ranking: Top 10 institutions and not in the Top 100 with respect the top 100; institutions sorted by their average number of citations: Low (0.1, 1.47), and High (1.77, 4.07) respect to Medium (1.48, 1.74); and Carnegie Selectivity Index, Inclusive and More selective with respect to Selective.

**Table S1. Number of papers, distinct authors and institutions, by institutional group.**

| Group | # papers | # authors | # institutions |
|---|---|---|---|
| **Carnegie Selectivity Index** | | | |
| more selective | 3264304 | 2280454 | 187 |
| selective | 1283167 | 833853 | 206 |
| inclusive | 262617 | 168731 | 196 |
| not indexed | 428957 | 168731 | 96 |
| **Average citations of institutions** | | | |
| High (1.77, 4.07) | 1826346 | 1215155 | 60 |
| Medium (1.48, 1.74) | 1815227 | 1189355 | 78 |
| Low (0.1, 1.47) | 1746770 | 1204592 | 547 |
| **US News & World ranking** | | | |
| Top 10 | 935931 | 579559 | 11 |
| Top 100 | 2517913 | 1684950 | 89 |
| Not in top | 1901865 | 1330966 | 584 |
| **Women and Minority Serving Institutions** | | | |
| HBCU | 35518 | 24829 | 62 |
| HSI | 278109 | 169469 | 127 |

| WC | 8710 | 5320 | 25 |

**Table S2. Parameters of the aggregated OLS model**

| variable | US News & World Report | Carnegie Selectivity Index | Average citations of institutions(discrete) | Average citations of institutions(continuous) |
|---|---|---|---|---|
| **topic normalized citations** | | | | |
| Intercept | 0.60*** (0.59, 0.62) | 0.53*** (0.51, 0.54) | 0.60*** (0.59, 0.61) | -0.16*** (-0.19, -0.14) |
| #authors | 0.03*** (0.03, 0.03) | 0.03*** (0.03, 0.03) | 0.03*** (0.03, 0.03) | 0.03*** (0.03, 0.03) |
| career age | 0.01*** (0.01, 0.01) | 0.01*** (0.01, 0.01) | 0.01*** (0.01, 0.01) | 0.01*** (0.01, 0.01) |
| Latinx men | -0.09*** (-0.12, -0.05) | -0.10*** (-0.13, -0.06) | -0.09*** (-0.13, -0.06) | -0.10*** (-0.13, -0.06) |
| Black men | -0.10*** (-0.15, -0.05) | -0.13*** (-0.18, -0.09) | -0.08*** (-0.13, -0.04) | -0.08** (-0.12, -0.03) |
| Asian men | 0.09*** (0.08, 0.10) | 0.09*** (0.07, 0.10) | 0.10*** (0.08, 0.11) | 0.09*** (0.08, 0.10) |
| White women | -0.07*** (-0.08, -0.06) | -0.08*** (-0.09, -0.06) | -0.07*** (-0.09, -0.06) | -0.06*** (-0.08, -0.05) |
| Latinx women | -0.13*** (-0.17, -0.09) | -0.14*** (-0.19, -0.10) | -0.14*** (-0.18, -0.09) | -0.13*** (-0.17, -0.09) |
| Black women | -0.17*** (-0.22, -0.11) | -0.19*** (-0.25, -0.14) | -0.17*** (-0.22, -0.11) | -0.15*** (-0.21, -0.09) |
| Asian women | 0.00 (-0.01, 0.02) | 0.00 (-0.01, 0.02) | 0.01 (-0.01, 0.02) | 0.00 (-0.01, 0.02) |
| usnr [top 10] | 0.57*** (0.56, 0.58) | | | |
| usnr [top 100] | 0.23*** (0.22, 0.24) | | | |
| selindex [more selective] | | 0.36*** (0.34, 0.38) | | |
| selindex [selective] | | 0.13*** (0.11, 0.15) | | |
| avg citations [high] | | | 0.48*** (0.47, 0.49) | |
| avg citations [medium] | | | 0.20*** (0.19, 0.21) | |
| avg citations | | | | 0.61*** (0.60, 0.62) |
| **topic normalized JIF** | | | | |
| Intercept | 0.78*** (0.78, 0.79) | 0.73*** (0.72, 0.73) | 0.78*** (0.77, 0.78) | 0.27*** (0.27, 0.28) |
| #authors | 0.01*** (0.01, 0.01) | 0.01*** (0.01, 0.01) | 0.01*** (0.01, 0.01) | 0.01*** (0.01, 0.01) |
| career age | 0.00*** (0.00, 0.00) | 0.00*** (0.00, 0.00) | 0.00*** (0.00, 0.00) | 0.00*** (0.00, 0.00) |
| Latinx men | -0.02*** (-0.04, -0.01) | -0.03*** (-0.05, -0.02) | -0.03*** (-0.04, -0.01) | -0.03*** (-0.04, -0.02) |
| Black men | -0.07*** (-0.09, -0.05) | -0.09*** (-0.11, -0.07) | -0.06*** (-0.08, -0.04) | -0.05*** (-0.07, -0.03) |
| Asian men | 0.06*** (0.05, 0.06) | 0.05*** (0.05, 0.06) | 0.06*** (0.05, 0.06) | 0.05*** (0.05, 0.06) |
| White women | -0.03*** (-0.04, -0.03) | -0.04*** (-0.04, -0.03) | -0.03*** (-0.04, -0.03) | -0.03*** (-0.03, -0.02) |
| Latinx women | -0.05*** (-0.07, -0.04) | -0.06*** (-0.08, -0.05) | -0.06*** (-0.08, -0.04) | -0.06*** (-0.07, -0.04) |
| Black women | -0.10*** (-0.12, -0.08) | -0.12*** (-0.14, -0.10) | -0.10*** (-0.12, -0.08) | -0.09*** (-0.11, -0.07) |
| Asian women | 0.03*** (0.02, 0.04) | 0.03*** (0.02, 0.04) | 0.03*** (0.03, 0.04) | 0.03*** (0.03, 0.04) |
| usnr [top 10] | 0.39*** (0.38, 0.39) | | | |
| usnr [top 100] | 0.16*** (0.16, 0.17) | | | |
| selindex [more selective] | | 0.25*** (0.24, 0.26) | | |
| selindex [selective] | | 0.09*** (0.08, 0.10) | | |

| variable | US News & World Report | Carnegie Selectivity Index | Average citations of institutions(discrete) | Average citations of institutions(continuous) |
|---|---|---|---|---|
| avg citations [high] | | | 0.33*** (0.33, 0.34) | |
| avg citations [medium] | | | 0.15*** (0.14, 0.15) | |
| avg citations | | | | 0.41*** (0.40, 0.41) |

**Table S3. Parameters of the disaggregated OLS models**

| Prestige group | variable | high prestige | medium prestige | low prestige |
|---|---|---|---|---|
| **topic normalized citations** | | | | |
| US News & World Report | Intercept | 1.24*** (1.21, 1.27) | 0.82*** (0.81, 0.84) | 0.48*** (0.46, 0.49) |
| | nb_auteur | 0.02*** (0.02, 0.02) | 0.03*** (0.03, 0.03) | 0.06*** (0.06, 0.06) |
| | career_age | 0.01*** (0.01, 0.01) | 0.01*** (0.01, 0.01) | 0.01*** (0.01, 0.01) |
| | hispanic_M | -0.08 (-0.21, 0.04) | -0.09*** (-0.14, -0.04) | -0.09*** (-0.13, -0.04) |
| | black_M | 0.01 (-0.17, 0.18) | -0.11** (-0.17, -0.04) | -0.11*** (-0.17, -0.04) |
| | asian_M | 0.07** (0.02, 0.12) | 0.08*** (0.06, 0.10) | 0.11*** (0.09, 0.13) |
| | white_F | -0.20*** (-0.25, -0.15) | -0.06*** (-0.08, -0.04) | -0.03** (-0.05, -0.01) |
| | hispanic_F | -0.36*** (-0.52, -0.20) | -0.13*** (-0.19, -0.07) | -0.06* (-0.12, -0.01) |
| | black_F | -0.20 (-0.40, 0.01) | -0.22*** (-0.30, -0.14) | -0.11** (-0.18, -0.03) |
| | asian_F | -0.04 (-0.10, 0.02) | 0.00 (-0.02, 0.03) | 0.02 (-0.01, 0.04) |
| Carnegie Selectivity Index | Intercept | 0.90*** (0.89, 0.92) | 0.53*** (0.51, 0.55) | 0.43*** (0.40, 0.46) |
| | nb_auteur | 0.03*** (0.03, 0.03) | 0.06*** (0.06, 0.06) | 0.06*** (0.05, 0.06) |
| | career_age | 0.01*** (0.01, 0.01) | 0.01*** (0.01, 0.01) | 0.01*** (0.01, 0.01) |
| | hispanic_M | -0.11*** (-0.15, -0.06) | -0.07* (-0.13, 0.00) | -0.18*** (-0.27, -0.09) |
| | black_M | -0.15*** (-0.21, -0.09) | -0.08 (-0.16, 0.00) | -0.17** (-0.29, -0.05) |
| | asian_M | 0.08*** (0.06, 0.10) | 0.08*** (0.05, 0.10) | 0.15*** (0.11, 0.19) |
| | white_F | -0.09*** (-0.11, -0.08) | -0.04** (-0.06, -0.02) | -0.04* (-0.08, 0.00) |
| | hispanic_F | -0.20*** (-0.25, -0.14) | -0.06 (-0.13, 0.01) | 0.00 (-0.10, 0.10) |
| | black_F | -0.23*** (-0.30, -0.16) | -0.14** (-0.24, -0.05) | -0.09 (-0.23, 0.05) |
| | asian_F | -0.01 (-0.03, 0.02) | 0.00 (-0.03, 0.03) | 0.06* (0.01, 0.11) |
| Average citations of institutions | Intercept | 1.08*** (1.06, 1.10) | 0.84*** (0.82, 0.86) | 0.49*** (0.47, 0.50) |
| | nb_auteur | 0.03*** (0.03, 0.04) | 0.02*** (0.02, 0.02) | 0.06*** (0.06, 0.06) |
| | career_age | 0.01*** (0.01, 0.01) | 0.01*** (0.01, 0.01) | 0.01*** (0.01, 0.01) |
| | hispanic_M | -0.07 (-0.15, 0.01) | -0.11*** (-0.17, -0.04) | -0.09*** (-0.13, -0.06) |
| | black_M | -0.07 (-0.18, 0.05) | -0.09* (-0.18, 0.00) | -0.07** (-0.12, -0.02) |
| | asian_M | 0.09*** (0.05, 0.12) | 0.07*** (0.04, 0.09) | 0.12*** (0.11, 0.14) |
| | white_F | -0.15*** (-0.18, -0.11) | -0.06*** (-0.09, -0.04) | -0.02** (-0.04, -0.01) |
| | hispanic_F | -0.32*** (-0.42, -0.22) | -0.08* (-0.16, -0.01) | -0.06* (-0.11, -0.01) |
| | black_F | -0.29*** (-0.42, -0.16) | -0.17** (-0.27, -0.07) | -0.10** (-0.16, -0.03) |
| | asian_F | -0.04* (-0.08, 0.00) | 0.00 (-0.03, 0.03) | 0.04*** (0.02, 0.07) |
| **topic normalized JIF** | | | | |
| US News & World Report | Intercept | 1.20*** (1.19, 1.22) | 0.94*** (0.94, 0.95) | 0.71*** (0.71, 0.72) |
| | nb_auteur | 0.01*** (0.01, 0.01) | 0.01*** (0.01, 0.01) | 0.03*** (0.03, 0.03) |
| | career_age | 0.00*** (0.00, 0.00) | 0.00*** (0.00, 0.00) | 0.00*** (0.00, 0.00) |
| | hispanic_M | 0.02 (-0.03, 0.07) | -0.03* (-0.05, -0.01) | -0.04*** (-0.06, -0.02) |
| | black_M | -0.14*** (-0.22, -0.07) | -0.04** (-0.07, -0.02) | -0.06*** (-0.09, -0.04) |
| | asian_M | 0.06*** (0.03, 0.08) | 0.06*** (0.05, 0.06) | 0.05*** (0.04, 0.06) |
| | white_F | -0.11*** (-0.13, -0.09) | -0.03*** (-0.03, -0.02) | -0.01* (-0.02, 0.00) |
| | hispanic_F | -0.14*** (-0.21, -0.07) | -0.04*** (-0.07, -0.02) | -0.04*** (-0.07, -0.02) |
| | black_F | -0.19*** (-0.28, -0.10) | -0.09*** (-0.12, -0.06) | -0.09*** (-0.12, -0.07) |
| | asian_F | 0.03* (0.00, 0.05) | 0.03*** (0.02, 0.04) | 0.03*** (0.02, 0.04) |
| Carnegie Selectivity Index | Intercept | 0.99*** (0.98, 0.99) | 0.75*** (0.74, 0.76) | 0.66*** (0.65, 0.68) |
| | nb_auteur | 0.01*** (0.01, 0.01) | 0.03*** (0.03, 0.03) | 0.03*** (0.02, 0.03) |
| | career_age | 0.00*** (0.00, 0.00) | 0.00*** (0.00, 0.00) | 0.00*** (0.00, 0.00) |
| | hispanic_M | -0.04*** (-0.05, -0.02) | -0.02* (-0.05, 0.00) | -0.05* (-0.09, -0.01) |
| | black_M | -0.10*** (-0.13, -0.08) | -0.07*** (-0.10, -0.04) | -0.05* (-0.11, 0.00) |
| | asian_M | 0.05*** (0.04, 0.06) | 0.04*** (0.03, 0.05) | 0.09*** (0.07, 0.10) |
| | white_F | -0.05*** (-0.05, -0.04) | -0.02*** (-0.03, -0.01) | 0.00 (-0.01, 0.02) |
| | hispanic_F | -0.08*** (-0.10, -0.06) | -0.03* (-0.06, 0.00) | -0.04* (-0.09, 0.00) |

| Prestige group | variable | high prestige | medium prestige | low prestige |
|---|---|---|---|---|
| | black_F | -0.13*** (-0.16, -0.10) | -0.10*** (-0.14, -0.07) | -0.10** (-0.16, -0.04) |
| | asian_F | 0.03*** (0.02, 0.04) | 0.03*** (0.02, 0.04) | 0.06*** (0.04, 0.08) |
| | Intercept | 1.13*** (1.12, 1.14) | 0.93*** (0.93, 0.94) | 0.70*** (0.70, 0.71) |
| | nb_auteur | 0.01*** (0.01, 0.01) | 0.01*** (0.01, 0.01) | 0.03*** (0.03, 0.03) |
| | career_age | 0.00*** (0.00, 0.00) | 0.00*** (0.00, 0.00) | 0.00*** (0.00, 0.00) |
| Average citations of institutions | hispanic_M | -0.01 (-0.04, 0.03) | -0.05*** (-0.07, -0.02) | -0.03*** (-0.05, -0.01) |
| | black_M | -0.11*** (-0.16, -0.06) | -0.03 (-0.06, 0.01) | -0.04** (-0.06, -0.01) |
| | asian_M | 0.06*** (0.05, 0.08) | 0.04*** (0.03, 0.05) | 0.06*** (0.06, 0.07) |
| | white_F | -0.08*** (-0.09, -0.07) | -0.02*** (-0.03, -0.01) | -0.01 (-0.01, 0.00) |
| | hispanic_F | -0.12*** (-0.17, -0.08) | -0.05** (-0.07, -0.02) | -0.03** (-0.05, -0.01) |
| | black_F | -0.17*** (-0.23, -0.12) | -0.09*** (-0.12, -0.05) | -0.07*** (-0.10, -0.04) |
| | asian_F | 0.01 (0.00, 0.03) | 0.03*** (0.02, 0.04) | 0.05*** (0.04, 0.06) |

**Table S4. Citation gap compared to White men, by identity and institutional group, normalized by topic.**

| Group | Black Men | Latinx Men | Asian Men | Asian Women | Latinx Women | Black Women | White Women |
|---|---|---|---|---|---|---|---|
| **US News & World Report** | | | | | | | |
| Top 10 | -0.48% | -3.19% | 0.26% | -9.20% | -19.84% | -15.54% | -15.15% |
| Top 100 | -2.32% | -4.56% | 2.76% | -4.78% | -10.83% | -11.15% | -8.07% |
| Not Top | -3.26% | -5.02% | 9.28% | -2.23% | -7.72% | -8.98% | -6.78% |
| **Carnegie Selectivity Index** | | | | | | | |
| More Selective | -2.90% | -4.77% | 2.85% | -5.49% | -14.10% | -12.83% | -10.21% |
| Selective | -2.53% | -3.78% | 4.71% | -4.27% | -8.44% | -10.56% | -7.96% |
| Inclusive | -5.68% | -11.77% | 17.04% | 4.02% | -2.66% | -8.69% | -6.69% |
| **Average citations of institutions** | | | | | | | |
| Low (0.1, 1.47) | -2.41% | -5.43% | 10.63% | 1.04% | -6.62% | -7.92% | -5.87% |
| Medium (1.48, 1.74) | -2.25% | -5.85% | 0.98% | -5.83% | -9.76% | -10.85% | -8.51% |
| High (1.77, 4.07) | -1.17% | -2.94% | 2.50% | -8.19% | -18.47% | -15.22% | -12.94% |

**Table S5. JIF gap compared to White men, by identity and institutional group, normalized by topic.**

| Group | Black Men | Latinx Men | Asian Men | Asian Women | Latinx Women | Black Women | White Women |
|---|---|---|---|---|---|---|---|
| **US News & World Report** | | | | | | | |
| Top 10 | -1.89% | 0.80% | 2.91% | -0.86% | -8.80% | -9.20% | -8.12% |
| Top 100 | -0.88% | -1.14% | 3.33% | 0.40% | -3.70% | -4.44% | -3.20% |
| Not Top | -1.56% | -1.88% | 4.41% | 2.25% | -3.04% | -3.77% | -1.92% |
| **Carnegie Selectivity Index** | | | | | | | |
| More Selective | -1.75% | -1.34% | 3.30% | 0.32% | -5.64% | -6.05% | -4.60% |
| Selective | -1.61% | -0.95% | 3.06% | 1.56% | -2.78% | -4.48% | -2.68% |
| Inclusive | -1.47% | -2.39% | 8.90% | 5.23% | -2.83% | -3.45% | -0.92% |
| **Average citations of institutions** | | | | | | | |
| Low (0.1, 1.47) | -0.95% | -1.31% | 5.70% | 3.56% | -1.94% | -3.11% | -1.67% |
| Medium (1.48, 1.74) | -0.65% | -2.30% | 2.14% | 0.33% | -4.03% | -4.35% | -3.14% |
| High (1.77, 4.07) | -1.50% | 0.10% | 3.83% | -1.04% | -7.79% | -7.99% | -6.66% |

**Table S6. Citation gap compared to White men, by identity and institutional group, normalized by field.**

| Group | Black Men | Latinx Men | Asian Men | Asian Women | Latinx Women | Black Women | White Women |
|---|---|---|---|---|---|---|---|
| **US News & World Report** | | | | | | | |
| Top 10 | -1.17% | -3.33% | 0.92% | -9.80% | -22.22% | -17.87% | -17.50% |
| Top 100 | -2.30% | -4.60% | 3.92% | -5.45% | -14.52% | -14.02% | -10.97% |
| Not Top | -3.24% | -4.75% | 11.17% | -2.52% | -8.51% | -10.33% | -7.99% |
| **Carnegie Selectivity Index** | | | | | | | |
| More Selective | -3.01% | -4.81% | 3.87% | -6.31% | -17.13% | -15.66% | -12.86% |
| Selective | -2.53% | -3.72% | 7.09% | -3.59% | -10.16% | -11.46% | -9.23% |
| Inclusive | -6.28% | -11.28% | 15.71% | 1.51% | -2.84% | -9.90% | -8.56% |
| **Average citations of institutions** | | | | | | | |
| Low (0.1, 1.47) | -2.56% | -4.93% | 12.53% | 1.08% | -7.79% | -9.39% | -7.00% |
| Medium (1.48, 1.74) | -1.94% | -5.78% | 1.71% | -6.55% | -13.26% | -13.68% | -11.51% |
| High (1.77, 4.07) | -1.61% | -3.29% | 3.79% | -8.99% | -21.48% | -17.85% | -15.61% |

**Table S7. JIF gap compared to White men, by identity and institutional group, normalized by field.**

| Group | Black Men | Latinx Men | Asian Men | Asian Women | Latinx Women | Black Women | White Women |
|---|---|---|---|---|---|---|---|
| **US News & World Report** | | | | | | | |
| Top 10 | -1.83% | 0.90% | 4.43% | -0.84% | -10.03% | -10.60% | -9.55% |
| Top 100 | -1.04% | -0.76% | 5.91% | 1.15% | -5.53% | -6.12% | -5.02% |
| Not Top | -1.46% | -1.02% | 7.13% | 3.42% | -3.97% | -5.04% | -3.11% |
| **Carnegie Selectivity Index** | | | | | | | |
| More Selective | -1.73% | -0.91% | 5.63% | 0.84% | -7.24% | -7.62% | -6.28% |
| Selective | -1.63% | -0.52% | 6.13% | 3.18% | -3.78% | -5.47% | -3.75% |
| Inclusive | -1.88% | -1.34% | 9.60% | 4.74% | -3.60% | -5.56% | -2.60% |
| **Average citations of institutions** | | | | | | | |
| Low (0.1, 1.47) | -1.00% | -0.14% | 8.61% | 4.93% | -3.10% | -4.79% | -3.03% |
| Medium (1.48, 1.74) | -0.76% | -1.93% | 4.74% | 1.06% | -5.77% | -5.71% | -4.85% |
| High (1.77, 4.07) | -1.52% | 0.05% | 5.66% | -0.88% | -9.22% | -9.49% | -8.27% |